\documentclass[11pt,a4paper]{article}
\pdfoutput=1
\usepackage[pdftex]{graphics}
\usepackage{jheppub}
\usepackage{amsmath,amssymb,amsfonts}
\usepackage{enumitem}
\usepackage{multirow}
\usepackage{array,booktabs}
\usepackage{slashed}
\usepackage[utf8]{inputenc}

\title{Poincaré invariance of spinning binary dynamics 
in the post-Minkowskian Hamiltonian approach}

\author[a]{Hojin Lee}
\author[a,b,c,d]{Sangmin Lee}

\affiliation[a]{Department of Physics and Astronomy, Seoul National University, \\ 1 Gwanak-ro, Gwanak-gu, Seoul 08826, Korea}
\affiliation[b]{Center for Theoretical Physics, Seoul National University, \\ 1 Gwanak-ro, Gwanak-gu, Seoul 08826, Korea}
\affiliation[c]{College of Liberal Studies, Seoul National University, \\ 1 Gwanak-ro, Gwanak-gu, Seoul 08826, Korea}
\affiliation[d]{School of Physics, Korea Institute for Advanced Study, \\ 85 Hoegi-ro, Dongdaemun-gu, Seoul 02455, Korea}

\emailAdd{zet4gra9er@snu.ac.kr, sangmin@snu.ac.kr}

\abstract{We initiate the construction of the global Poincaré algebra generators 
in the context of the post-Minkowskian Hamiltonian formulation of gravitating binary dynamics in isotropic coordinates that is partly inspired by scattering amplitudes. 
At the first post-Minkowskian (1PM) order, we write down the Hamiltonian in a form valid in an arbitrary inertial frame. 
Then we construct the boost generator at the same order which uniquely solves all the equations required by the Poincaré algebra. Our results are linear in Newton's constant but exact in velocities and spins, including all spin-multipole moments. We also construct explicitly the canonical transformations that prove 
the equivalence between our new generators and the corresponding generators in the ADM coordinates up to the second post-Newtonian (2PN) order.}

\begin{document}

\maketitle

\section{Introduction}

A two-body problem in general relativity is often addressed by an effective theory. 
The fundamental theory is the Einstein-Hilbert action coupled to the two bodies. 
The effective theory is obtained by ``integrating out" the gravitational field 
mediating the interaction. 
To be specific, we restrict our attention to the Hamiltonian formulation and focus on the conservative sector where the gravitational radiation is not yet taken into account. 

There are two widely used perturbative expansions to build up the effective theory. One is the post-Newtonian (PN) expansion 
in powers of $v^2 \sim GM/r$. The other is the post-Minkowskian (PM) expansion in powers of $GM/r$ 
with no assumption on $v^2$. 
See {\it e.g.} \cite{Blanchet:2013haa,Porto:2016pyg,Schafer:2018kuf,Levi:2018nxp,Bjerrum-Bohr:2022blt,Kosower:2022yvp,Buonanno:2022pgc,Goldberger:2022ebt} for comprehensive reviews and further references. 

Regardless of the perturbative scheme, the two-body effective theory 
is expected to be covariant under the global Poincaré symmetry 
of the underlying asymptotically flat spacetime. 
In the traditional ADM formalism, the structure of the Poincaré generators was elucidated in \cite{Regge:1974zd}. 
The generators are of great practical values too.
As shown in \cite{Damour:2000kk}, whenever one reaches a new order of the (PN or PM) perturbative Hamiltonian, 
constructing the Poincaré generators to the same order provides a stringent consistency check 
and can sometimes fix undetermined coefficients. 
The boost generator of a spinning binary was first constructed in \cite{Damour:2007nc} up to the next-to-leading order (NLO) spin-orbit (SO) coupling. 
Its extension to higher order can be found {\it e.g.} in \cite{Steinhoff:2008zr,Hergt:2008jn,Hartung:2013dza,Levi:2016ofk}. 
Recently, the construction was completed up to the overwhelming 5PN order 
in \cite{Levi:2022dqm,Levi:2022rrq}. 

A recent trend in this subject area is that ideas and tools from (quantum) scattering amplitudes 
are being used to produce new results in the (classical) PM Hamiltonian; 
see {\it e.g.} \cite{Bjerrum-Bohr:2022blt,Kosower:2022yvp,Buonanno:2022pgc} for reviews. 
Scattering amplitudes are well known to be Lorentz scalars. 
As such, one may expect that the PM Hamiltonian based on amplitudes 
should be well-suited for the discussion of the global Poincaré symmetry. 
Curiously, however, there has been little to no attempts 
to construct the Poincaré generators in the amplitude-based PM context. 
The main goal of the present paper is to fill this gap.

For simplicity, let us begin with a non-spinning binary system. 
The first order post-Minkowskian (1PM) Hamiltonian in the center of momentum (COM) frame 
is \cite{Cheung:2018wkq} 
\begin{align}
\begin{split}
        &H^{[1]}|_\mathrm{COM} = - \frac{Gm_1^2 m_2^2(2\gamma^2 -1)}{E_1 E_2 r } \,, 
        \quad \gamma = -\frac{p_1 \cdot p_2}{m_1m_2} = \frac{1}{m_1 m_2} (E_1 E_2 - \vec{p}_1 \cdot \vec{p}_2) \,.
\end{split}
    \label{H-1PM-bare}
\end{align}
Our metric convention is $(-+++)$. Momentum 4-vectors 
are denoted by $p_a = (E_a,\vec{p}_a)$ $(a=1,2)$ in the $c=1$ unit.
The COM frame is defined by  $\vec{p}_1 + \vec{p}_2 = 0$. 
Once we choose an inertial frame, 
most equations are written in the 3d vector notation.
 Abbreviations used here and to be used frequently throughout this paper include
\begin{align}
\begin{split}
   &\vec{r} = \vec{x}_1 - \vec{x}_2 \,, \quad 
   r = |\vec{r}| \,, \quad 
   \hat{n} = \frac{\vec{r}}{r} \,, 
    \\
    &E_a = \sqrt{\vec{p}_a^2 + m_a^2} \,, \quad 
    E = E_1 +E_2\,, \quad 
    z_a = \frac{E_a}{E_1+E_2}   \,, \quad 
    \vec{u}_a = \frac{\vec{p}_a}{E_a}  \,.
\end{split}
\end{align} 
In \eqref{H-1PM-bare}, the superscript of $H^{[1]}$ denotes the 1PM order that is linear in Newton's constant $G$ but 
exact in $\vec{v}=\vec{p}/m$. 
Among the Poincaré generators, the most interesting and delicate is the boost generator, 
as we will review shortly.
A boost generator connects distinct inertial frames. 
In order to construct the boost generator associated with the Hamiltonian \eqref{H-1PM-bare}, 
we have to generalize it to a form valid in an arbitrary inertial frame. 
The desired generalization will be the first new result of this paper in section~\ref{sec:ns-1pm-H}. 
\footnote{In the ADM formulation, the 1PM Hamiltonian in a closed form was first derived by Ledvinka et al \cite{Ledvinka:2008tk}. 
It was shown to be consistent with PN computations up to the 4PN order in \cite{Jaranowski:2012eb}.}

Given the general form of the Hamiltonian,  
the closure of the Poincaré algebra uniquely determines the boost generator, 
as we will show in section~\ref{sec:ns-1pm-G}. 
The computation is based on a mild ansatz to be explained later in this section.  
In short, our ansatz is motivated by a similar ansatz in \cite{Damour:2000kk} 
and the fact that the boost generator defines a sort of ``center coordinate" of the binary system. 

In section~\ref{sec:spin-1pm}, we repeat the same exercise for a spinning binary system. 
Our work for spinning binaries remains at the 1PM order but is exact not only in velocities but also in spins. 
Building upon the general spin-multipole expansion of \cite{Levi:2014gsa,Levi:2015msa} 
and its on-shell interpretation of \cite{Arkani-Hamed:2017jhn} in terms of the massive spinor helicity variables, the complete spinning 1PM Hamiltonian was first obtained in the COM frame in \cite{Chung:2020rrz}. 
Again, we generalize the Hamiltonian to a form valid in an arbitrary frame 
and then construct the boost generator. 
A proper understanding of the ``spin-gauge symmetry" in the Hamiltonian framework 
plays a critical role throughout our computations. See \cite{Steinhoff:2015ksa,Kim:2021rda} 
and appendix~\ref{sec:spin-gauge-symm} of this paper 
for a modern treatment of the spin-gauge symmetry.

PN and PM computations should agree whenever they overlap. 
In appendix~\ref{sec:more-PN}, we spell out explicitly the spin-orbit coupling part of our PM Hamiltonian to facilitate the comparison. 
In appendix~\ref{sec:PN-check}, we verify explicitly that our new results are consistent with 
PN results in the literature, as well as the 1PM Hamiltonian of \cite{Ledvinka:2008tk}. 
In the non-spinning case, 
we establish the equivalence between our Hamiltonian and the ADM Hamiltonian 
at the intersection of the 1PM and 2PN expansions. 
In the spinning case, 
we establish the equivalence up to the NLO SO coupling linear in $G$. 
The boost generators are also compared and shown to agree. 
In practice, showing the equivalence means
finding an explicit form of the generator of a canonical transformation 
which accounts for the difference between the two coordinate choices. 

Our work naturally suggests many directions for generalizations and applications. Some of them will be discussed in section~\ref{sec:discussion}.

\subsection{Poincaré algebra} \label{sec:ansatz}

In this technical appendix to the Introduction, we briefly review the general structure of the Poincaré algebra generators following \cite{Damour:2000kk}, 
and motivate our ansatz that will play a key role in the main text. 

For a 2-body dynamics with spin, the translation and rotation generators are simply 
\begin{align}
    \vec{P} = \vec{p}_1 + \vec{p}_2 \,,
    \quad 
    \vec{J} = \sum_{a=1}^2 \left( \vec{x}_a \times \vec{p}_a + \vec{S}_a \right) \,.
    \label{P-J}
\end{align}
Our convention for the Poisson bracket is such that 
\begin{align}
    \{ x_{a,i}, p_{b,j} \} = \delta_{ab} \delta_{ij} \,,
    \quad 
    \{ S_{a,i} , S_{b,j} \} = \delta_{ab} \epsilon_{ijk} S_{a,k} \,.
    \label{3d-Poisson}
\end{align}
It is customary to trade the boost generator $\vec{K}$ for $\vec{G}$ 
through the relation $\vec{K} = \vec{G} - t \vec{P}$. 
The algebra then reads 
\begin{align}
& \left\{P_i, P_j\right\}= 0 \,, \quad \qquad \left\{P_i, H\right\}=0 \,, \quad \qquad \left\{J_i, H\right\}=0 \,, 
\label{PA-1}
\\
&\left\{J_i, J_j\right\}= \epsilon_{i j k} J_k \,, \quad\left\{J_i, P_j\right\}=\epsilon_{i j k} P_k \,, \quad\left\{J_i, G_j\right\}=\epsilon_{i j k} G_k\,, 
\label{PA-2}
\\
& \left\{G_i, P_j\right\}=\delta_{i j} H \,, \quad\left\{G_i, H\right\}=P_i \,, \quad\qquad \left\{G_i, G_j\right\}=-\epsilon_{i j k} J_k \,.
\label{PA-3}
\end{align}
Translation and rotation invariance of $H$ in \eqref{PA-1} dictates that $H$ is a 3d scalar 
function that depend on inner products among $\vec{p}_1$, $\vec{p}_2$, $\vec{S}_1$, $\vec{S}_2$ and $\vec{r} = \vec{x}_1 - \vec{x}_2$. 
The third relation in \eqref{PA-2} means that $\vec{G}$ is a 3d-vector. 

The non-trivial checks lie in \eqref{PA-3}. When $H$ has been computed up to some order in a perturbation series,  
we should construct, to a commensurate order, $\vec{G}$ satisfying all three conditions in \eqref{PA-3}. 
Let us call the three conditions ``$H$/$P$/$J$-conditions", respectively. 
\textcolor{blue}{
The construction of $\vec{G}$ boils down to writing down an ansatz and solving the $H$/$P$/$J$-conditions.
}

The $H/P$-conditions give some hints on how to write down an ansatz for $\vec{G}$ at each order. 
Since $\vec{P}$ and $H$ commutes with each other, 
we can rewrite the $H/P$-conditions as
\begin{align}
    \vec{X} \equiv \frac{\vec{G}}{H}  
    \quad 
    \Longrightarrow
    \quad 
    \{ X_i , P_j \} = \delta_{ij} 
    \,,
    \quad 
    \{ \vec{X} , H \} = \frac{\vec{P}}{H} \,.
\end{align}
The first relation on the RHS states that $\vec{X}$ acts like a position coordinate canonically conjugate to $\vec{P}$. 
The second relation states that the velocity of $\vec{X}$ stays constant:
\begin{align}
    \frac{d\vec{X}}{dt} = \{ \vec{X} , H \} = \frac{\vec{P}}{H} \,.  
\end{align}
These observations are well known and appealing, but in practice when we rely on a perturbation theory, the non-perturbative relations may not be directly useful. 
With hindsight, we suggest that we introduce the ``coordinate" vector $\vec{X}$ order by order. 
Specifically, for the PM expansion, 
\begin{align}
    H = H^{[0]} +  H^{[1]} +  H^{[2]} + \cdots \,,
    \qquad 
    \vec{G} = \vec{G}^{[0]} +  \vec{G}^{[1]} +  \vec{G}^{[2]} + \cdots \,,
\end{align}
\textcolor{blue}{in the non-spinning case, the most general ansatz compatible with symmetries is }
%
\begin{align}
     \vec{G}^{[n]} =  H^{[n]} \vec{X}^{[n]} \,,
    \quad 
    \vec{X}^{[n]} = \alpha^{[n]}_1 \vec{x}_1 + \alpha^{[n]}_2 \vec{x}_2 +   \beta^{[n]}_1 \vec{p}_1 + \beta^{[n]}_2 \vec{p}_2 \,,
    \label{G-X-ansatz}
\end{align}
\textcolor{blue}{where $\alpha^{[n]}_{1,2}$ and $\beta^{[n]}_{1,2}$ are functions of $\vec{p}_1$, $\vec{p}_2$ and $\vec{r}$. 
$\alpha^{[n]}_{1,2}$ are dimensionless and $\beta^{[n]}_{1,2}$ has the dimension of length/mass.}

\textcolor{blue}{
While it is certainly desirable to begin with the most general ansatz and constrain the coefficient functions 
by imposing the $H$/$P$/$J$-conditions, 
we take a shortcut by \emph{assuming the uniqueness}\footnote{\textcolor{blue}{Suppose $\vec{G}$ is not unique. Let $\vec{G}$ and $\vec{G}'$ be two distinct vectors satisfying all $H$/$P$/$J$-conditions. Then $\Delta \vec{G} = \vec{G}' - \vec{G}$ would give a new vector conserved charge: $d(\Delta\vec{G})/dt = \{ \Delta \vec{G} , H \} =0$. We find such an extra conserved charge extremely unlikely, although we cannot exclude the possibility completely rigorously.}} 
of $\vec{G}$ and working with a simpler ansatz inclusive enough to admit a solution. 
Remarkably, this work and a sequel \cite{Lee:2023zuu} find that it is safe to set }
\begin{align}
     \beta^{[n]}_{1,2} = 0  \,,
     \quad n = 0, 1, 2 \,.
\label{vanishing-beta}
\end{align}
\textcolor{blue}{
The ansatz used in the ADM coordinates \cite{Damour:2000kk} contains terms proportional to $\vec{p}_{1,2}$. 
But, in the PM isotropic coordinates used in this paper and \cite{Lee:2023zuu}, 
the vanishing of $\beta^{[n]}_{1,2}$ is compatible with the Poincar\'e algebra. 
%
%
A similar statement can be made on the $\vec{r}$-independence of $\alpha^{[1]}_{1,2}$. 
It would be nice to give a more intuitive explanation for this drastic simplification
and to see if the vanishing of $\beta^{[n]}_{1,2}$ persists for $n\ge 3$. 
We leave these questions for future work. 
}

At 0PM (free theory), the coefficients $\alpha_{1,2}^{[0]}$ are easy to read off:
\begin{align}
    \vec{G}^{[0]} = E_1 \vec{x}_1 + E_2 \vec{x}_2 
    \quad 
    \Longrightarrow
    \quad 
    \vec{X}^{[0]} = \left( \frac{E_1}{E_1 + E_2} \right) \vec{x}_1 + \left( \frac{E_2}{E_1 + E_2} \right) \vec{x}_2 \,.
\end{align}
At each order, the $H$-condition requires that 
\begin{align}
   \alpha^{[n]}_1 +  \alpha^{[n]}_2 = 1 \,. 
\end{align}
The $P$ and $J$ conditions are non-trivial because $H$ and $\vec{G}$ receive perturbative corrections. 

In the main text of this paper, we will apply the ansatz \eqref{G-X-ansatz} and its extension to the spinning case 
to the 1PM order and successfully solve the $P$ and $J$ conditions, thereby completing the 
construction of $\vec{G}^{[1]}$.

\section{Non-spinning 1PM} \label{sec:ns-1pm}

\subsection{Hamiltonian} \label{sec:ns-1pm-H}

We recall the process of turning a tree-level amplitude to a potential. 
In the 2-to-2 elastic scattering, the momentum transfer 4-vector $(\hbar q)$ is defined such that 
\begin{align}
\begin{split}
    p_1^\mathrm{(in)} = p_1 - \hbar q/2 \,, 
    &\quad 
    p_2^\mathrm{(in)} = p_2 + \hbar q/2 \,, 
    \\
    p_1^\mathrm{(out)}  = p_1 + \hbar q/2 \,, 
    &\quad 
    p_2^\mathrm{(out)} = p_2 - \hbar q/2 \,, 
\end{split}
\end{align}
The $\hbar$-scaling of $q$ (and no such scaling for $p_a$) stems from the fact that the classical limit 
of a scattering process is a large angular momentum limit in units of $\hbar$. 
Schematically, for a typical asymptotic momentum $\vec{p}$ and impact parameter $\vec{b}$, we have 
\begin{align}
    |\vec{b}\times\vec{p} | \gg \hbar \,,
    \quad 
    |\vec{b}| \sim |\vec{q}|^{-1} 
    \quad 
    \Longrightarrow 
    \quad 
    |\vec{p}| \gg \hbar |\vec{q}| \,.
\end{align}
The second relation reflects the fact (to be explained shortly) that $\vec{q}$ is the Fourier-dual of the relative position 
between the two bodies.

Requiring the mass-shell condition for all external momenta, 
we find two conditions: 
\begin{align}
    (p_1 +p_2 )\cdot q = 0 \,,
    \quad 
    (p_1 - p_2 )\cdot q = 0  \,.
\label{which-q}
\end{align}
The space of $q$ satisfying \emph{both} conditions 
is 2-dimensional. But, as is well known 
for the Newton or the Coulomb potential, we have to perform a 3d Fourier transform to obtain the potential. 
There is no real tension here; the scattering amplitude is on-shell whereas the Hamiltonian is off-shell. 
The standard procedure is to impose only the first condition in \eqref{which-q} and relax the second one.

For a given inertial frame (call it the lab frame), 
we can solve the condition for $q^0$: 
\begin{align}
    q^0 = \frac{(\vec{p}_1 + \vec{p}_2) \cdot\vec{q}}{E_1 +E_2} 
    \equiv \vec{u}_c \cdot \vec{q} \,. 
    \label{q0-uc}
\end{align}
Here, $\vec{u}_c$ is the velocity of the 
center of momentum degree of freedom 
as observed by the lab frame. 
In the COM frame, the famous Fourier transform 
that turns a tree amplitude to a potential is 
\begin{align}
     4\pi \int_{\vec{q}}   \,  \, \frac{e^{i\vec{q}\cdot\vec{r}}}{\vec{q}^2} = \frac{1}{r} \,, 
     \qquad 
     \int_{\vec{q}} \equiv  \int \frac{d^3\vec{q}}{(2\pi)^3} \,.
\end{align}
To work in a non-COM frame, we simply take 
\begin{align}
    \mathcal{I} =  4\pi \int_{\vec{q}} \,  \, \frac{e^{i\vec{q}\cdot\vec{r}}}{\vec{q}^2- (q^0)^2} = 
     4\pi \int_{\vec{q}}  \,  \, \frac{e^{i\vec{q}\cdot\vec{r}}}{\vec{q}^2- (\vec{u}_c\cdot \vec{q})^2 } \,.
\end{align}
(This integral was considered earlier in \cite{Jones:2022aji}.) 
One can perform the integral by noting that 
\begin{align}
    \vec{q}^2 - (\vec{u}_c\cdot \vec{q})^2 = q^i A_{ij} q^j \,,
\end{align}
where $A_{ij}$ is an effective ``metric" which depends on $\vec{p}_{1,2}$ but is constant on the $\vec{q}$-space. 
Since $\vec{u}_c^2 < 1$, the metric $A_{ij}$ is non-generate. 
In the end, the integral simply gives 
\begin{align}
     \mathcal{I} = \frac{1}{r\sqrt{1- (\vec{u}_{c})^2_\perp} } \equiv \frac{\gamma_c}{r} \,,
\end{align}
where we introduced a notation for the inner product on the plane transverse to $\hat{n}$:
\begin{align}
    (\vec{A}\cdot \vec{B})_\perp \equiv \vec{A} \cdot \vec{B} - (\vec{A} \cdot \hat{n}) ( \hat{n} \cdot \vec{B}) \,,
    \quad 
    (\vec{u}_{c})^2_\perp \equiv (\vec{u}_c \cdot \vec{u}_c)_\perp \,.
\end{align}
We may say that the potential acquires a Lorentz factor 
associated with the transverse part of $\vec{u}_c$. 
In conclusion, the 1PM Hamiltonian in the lab frame is given by 
\begin{align}
    H^{[1]} = \gamma_c \, H^{[1]}_\mathrm{bare}  = \gamma_c \left[  - \frac{Gm_1^2 m_2^2(2\gamma^2 -1)}{E_1 E_2 r}  \right]\,.
    \label{H-1PM-full}
\end{align} 
The $H^{[1]}_\mathrm{bare}$ part looks identical to the COM Hamiltonian $H^{[1]}_\mathrm{COM}$ \eqref{H-1PM-bare}, but it is no longer subject to 
the COM condition $\vec{p}_1 + \vec{p}_2 =0$. In other words, $H^{[1]}_\mathrm{bare}$ is a function of two independent variables 
$(\vec{p}_1, \vec{p}_2)$ 
whereas $H^{[1]}_\mathrm{COM}$ is a function of one variable ($\vec{p} = \vec{p}_1 = - \vec{p}_2$). 

\subsection{Boost} \label{sec:ns-1pm-G}

Our next goal is to construct the 1PM boost generator $\vec{G}^{[1]}$ compatible with our new Hamiltonian $H^{[1]}$  \eqref{H-1PM-full}, 
starting from the \textcolor{blue}{ansatz \eqref{G-X-ansatz} with \eqref{vanishing-beta}} and 
solving the $P$/$J$ conditions:
\begin{align}
    \vec{G}^{[1]} =  H^{[1]} \vec{X}^{[1]} 
    = H^{[1]} (\alpha^{[1]}_1 \vec{x}_1 + \alpha^{[1]}_2 \vec{x}_2) \,, 
    \quad 
    \alpha^{[1]}_1  + \alpha^{[1]}_2 = 1 \,.
    \label{G1-ansatz}
\end{align}
To avoid clutter, we will omit the superscript from $\alpha^{[1]}_a$ in what follows.

\paragraph{P first}

At first sight, it is not clear whether we should try to solve the $P$-condition first or the $J$-condition first. 
Let us explore both possibilities, 
starting from the $P$-condition:
\begin{align}
\{ \vec{G}^{[0]} , H^{[1]} \} + \{ \vec{G}^{[1]} , H^{[0]} \} = 0 \,.
\label{P-cond-1PM}
\end{align}
The first term in \eqref{P-cond-1PM} can be computed explicitly: 
\begin{align} \label{G0-H1}
\begin{split}
     \{\vec{G}^{[0]}, H^{[1]}\} &=  -\left( \vec{u}_1 + \vec{u}_2 \right) H^{[1]}+ (\vec{u}_c - \gamma_c^2 (\hat{n}\cdot\vec{u}_c)\hat{n})  H^{[1]}
     \\
     &\quad 
   + \left(\frac{\vec{x}_1}{r}  (\hat{n}\cdot \vec{u}_1 ) -\frac{\vec{x}_2}{r} (\hat{n}\cdot \vec{u}_2 ) \right) H^{[1]}
   \\
   &\quad 
   + \gamma_c^2 (\hat{n}\cdot\vec{u}_c) \left( \frac{\vec{x}_1}{r} (\vec{u}_c\cdot \vec{u}_1)_\perp - \frac{\vec{x}_2}{r} (\vec{u}_c\cdot \vec{u}_2)_\perp \right) H^{[1]}\,.
\end{split}
\end{align}
The Lorentz-scalar $(2\gamma^2-1)$ commutes with $\vec{G}^{[0]}$ as a consequence of 
\begin{align}
    \{ \vec{G}^{[0]} , \gamma \} = 0 \,.
\end{align}
From here on, any function of $\gamma$ will be treated as a constant. 

As for the second term in \eqref{P-cond-1PM}, the ansatz \eqref{G1-ansatz} give 
\begin{align} \label{G1-H0-x}
\begin{split}
    \{ M_1 \vec{x}_1 + M_2 \vec{x}_2 , H^{[0]} \}  &= M_1 \vec{u}_1 + M_2 \vec{u}_2 
    + \{ M_1  , H^{[0]} \}\vec{x}_1 + \{ M_2  , H^{[0]} \}\vec{x}_2 \,, 
\end{split}
\end{align}
where $M_a = H^{[1]} \alpha_a$ $(a=1,2)$. 
The next step is to match the coefficients of the four vectors $\vec{u}_1$, $\vec{u}_2$, $\vec{x}_1$, $\vec{x}_2$ from \eqref{P-cond-1PM}:
\begin{align}
\begin{split}
      M_1  - \frac{E_2}{E_1+E_2} H^{[1]}&= 0\,,
    \\
      M_2  - \frac{E_1}{E_1+E_2} H^{[1]}&= 0\,,
\\
      \{M_1, H^{[0]} \} - \frac{1}{r} \left[ \gamma_c^2 (\hat{n}\cdot \vec{u}_c) [1- (\vec{u}_c\cdot \vec{u}_1)_\perp ] - (\hat{n}\cdot \vec{u}_1) \right]H^{[1]}&= 0 \,,
      \\
      \{M_2, H^{[0]} \} + \frac{1}{r} \left[ \gamma_c^2 (\hat{n}\cdot \vec{u}_c) [1- (\vec{u}_c\cdot \vec{u}_2)_\perp ] - (\hat{n}\cdot \vec{u}_2) \right]H^{[1]}&= 0 \,.
\end{split}
\label{G1-uxeq}
\end{align}
Remarkably, the first two conditions already fix 
the sought-for coefficients $\alpha_{1,2}$ uniquely, 
\begin{align}
    \alpha_1 = z_2 \,,
    \quad 
    \alpha_2 = z_1 \,,
    \quad 
    z_a \equiv \frac{E_a}{E_1+E_2} \,.
    \label{alpha-sol}
\end{align}
Moreover, a bit of algebra shows that this solution also passes the other two, seemingly over-constraining, conditions. To recap, we have shown with mild effort that 
\begin{align}
    \vec{G}^{[1]} = H^{[1]}\vec{X}^{[1]} \,,
    \quad
    \vec{X}^{[1]} = z_2 \vec{x}_1 + z_1 \vec{x}_2  \,.
    \label{G1-nospin-final}
\end{align}

Once the $P$-condition is established, 
it takes not too much work to verify the $J$-condition. We rewrite the $J$-condition with a short-hand notation as
\begin{align}
    \{\vec{G}^{[0]} , \vec{G}^{[1]} \}_{\times} = 0 \,,
    \qquad 
    \{ \vec{A} , \vec{B} \}_{\times}|_i \equiv \epsilon_{ijk} \{ A_j , B_k \} \,.
\end{align}
We can simplify the $J$-condition using the $P$-condition as follows.
\begin{align}
\begin{split}
       \{\vec{G}^{[0]} , \vec{G}^{[1]} \}_{\times} &= \{\vec{G}^{[0]} , H^{[1]}\vec{X}  \}_{\times}
       \\
       &= \{\vec{G}^{[0]} , H^{[1]}  \}\times \vec{X} + H^{[1]}\{\vec{G}^{[0]} ,  \vec{X}  \}_{\times}
       \\
       &= - \{\vec{G}^{(1)} , H^{[0]}   \}\times \vec{X} + H^{[1]}\{\vec{G}^{[0]} ,  \vec{X}  \}_{\times}
       \\
       &= - \{H^{[1]}\vec{X} , H^{[0]}   \}\times \vec{X} + H^{[1]}\{\vec{G}^{[0]} ,  \vec{X}  \}_{\times}
       \\
       &= H^{[1]}\left[ \{H^{[0]} , \vec{X} \} \times \vec{X} + \{\vec{G}^{[0]} ,  \vec{X}  \}_{\times} \right] \,. 
\end{split}
\label{J-1PM-new}
\end{align}
The $\{H^{[1]}, H^{[0]} \}$ term drops out because $\vec{X} \times \vec{X} = 0$. 
Evaluating the brackets, we find 
\begin{align}
\begin{split}
    &\qquad \{H^{[0]} , \vec{X} \} \times \vec{X} + \{\vec{G}^{[0]} ,  \vec{X}  \}_{\times} 
    \\
    &= 
    \left[ \alpha_1^2 - \alpha_1 - E_1 \partial_{E_1} \alpha_1 \right] (\vec{x}_1 \times \vec{u}_1) + \left[ \alpha_1 \alpha_2 - E_2 \partial_{E_2} \alpha_1 \right]  (\vec{x}_1 \times \vec{u}_2 )
    \\
    &\quad + \left[ \alpha_2^2 - \alpha_2 - E_2 \partial_{E_2} \alpha_2 \right] (\vec{x}_2 \times \vec{u}_2) + \left[ \alpha_1 \alpha_2 - E_1 \partial_{E_1} \alpha_2 \right]  (\vec{x}_2 \times \vec{u}_1 ) \,. 
    \label{alpha12-eqs}
\end{split}
\end{align}
So, the $J$-condition gives four partial differential equations for the two functions $\alpha_{1,2}$. 
Since $\alpha_1 + \alpha_2 = 1$, we may focus on the two equations for $\alpha_1$. 
When we already have a candidate solution \eqref{alpha-sol}, 
it is trivial to check that they indeed satisfy the equations. 

\paragraph{J first and P-condition in q-space} 

We can try to solve the $J$-condition first and then verify the $P$-condition. 
But, to minimize extra work, we take \eqref{J-1PM-new} as our starting point, 
where it is \emph{assumed} that the $P$-condition holds. 
This may sound tautological, but as long as we confirm the $P$-condition 
after we solve the $J$-condition, there will be no logical flaw. 

Combining the two partial differential equations for $\alpha_1$ from the vanishing of \eqref{alpha12-eqs}, we can show that $\alpha_1$ depends only on the ratio $E_1/E_2$. 
Each equation can be solved up to an integration constant, giving
\begin{align}
    \alpha_1 = \frac{1}{1+ \kappa_1 (E_1/E_2)} = 1 - \frac{1}{1+\kappa_2 (E_2/E_1)} \,.
\end{align}
The exchange symmetry of the two particles sets $\kappa_1 = 1 = \kappa_2$, confirming that \eqref{alpha-sol} 
is the unique solution to the $J$-condition. 
We may now proceed to verify the $P$-condition.

In our derivation of the 1PM Hamiltonian \eqref{H-1PM-full}, 
the momentum space expression of the amplitude played a crucial role. 
It will become even more important when we add spins to the binary system in the next section. 
In that regard, it is instructive to reexamine the $P$-condition in the $\vec{q}$-space, 
where the ``Hamiltonian" is 
\begin{align}
    \widetilde{H}^{[1]} = \left( \frac{k}{E_1E_2} \right) \frac{e^{i\vec{q}\cdot \vec{r}}}{\vec{q}^2 - (q^0)^2} \,, 
    \quad 
    k = - 4\pi G m_1^2m_2^2 (2\gamma^2-1) \,,
    \quad 
    H^{[1]} = \int_{\vec{q}}  \widetilde{H}^{[1]} \,.
    \label{H1-q-space}
\end{align}
Recall that the $P$-condition can be written as 
\begin{align}
    \{ H^{[0]} \vec{X}^{[0]} , H^{[1]} \} + \{ H^{[1]} \vec{X}^{[1]} , H^{[0]} \} = 0 \,.
\end{align}
Rearranging a little bit, we obtain 
\begin{align}
    (\vec{X}^{[0]} - \vec{X}^{[1]})  \{ H^{[0]}  , H^{[1]} \} + H^{[0]}\{\vec{X}^{[0]} , H^{[1]} \} + H^{[1]}\{\vec{X}^{[1]} , H^{[0]} \} = 0 \,.
    \label{P-HXHX} 
\end{align}
In the $\vec{q}$-space, a short computation shows, 
\begin{align}
\begin{split}
        \{ H^{[0]} , \widetilde{H}^{[1]} \} &= -i\vec{q}\cdot ( \vec{u}_1 - \vec{u}_2 ) \widetilde{H}^{[1]} \,,
        \\
        \{ \vec{X}^{[0]} , \widetilde{H}^{[1]} \} &= \left[ -i\vec{q}\cdot (z_2\vec{u}_1 + z_1 \vec{u}_2) \vec{r} + \frac{2(\vec{u}_c\cdot\vec{q}) \left(\vec{q} - (\vec{u}_c\cdot\vec{q})\vec{u}_c \right)}{\vec{q}^2 - (q^0)^2}  \right] \frac{\widetilde{H}^{[1]}}{H^{[0]}} 
        \\
        &\qquad - (\vec{u}_1 +\vec{u}_2) \frac{\widetilde{H}^{[1]}}{H^{[0]}} \,,
        \\
        \{ \vec{X}^{[1]} , H^{[0]} \} &= z_2 \vec{u}_1 + z_1 \vec{u}_2 = (\vec{u}_1 +\vec{u}_2) - \vec{u}_c \,. 
\end{split}
\label{P-q-steps}
\end{align}
We combine these to find 
\begin{align}
\begin{split}
      &(\vec{X}^{[0]} - \vec{X}^{[1]})  \{ H^{[0]}  , 
      \widetilde{H}^{[1]} \} + H^{[0]}\{\vec{X}^{[0]} , \widetilde{H}^{[1]} \} + \widetilde{H}^{(1)}\{\vec{X}^{[1]} , H^{[0]} \} 
      \\
      &\hskip 4cm = \nabla_{\vec{q}} \left[ -(\vec{u}_c\cdot \vec{q})  \widetilde{H}^{[1]} \right]  \,.
\end{split}
\label{P-q-total}
\end{align}
The integration of the RHS over the $\vec{q}$-space can be turned to a surface integral. 
Because of the extra factor $(\vec{u}_c \cdot\vec{q})$ in \eqref{P-q-total}, 
the integrand of the surface integral may appear to grow linearly in $|\vec{q}|$. 
But, the rapid oscillation of the $e^{i\vec{q}\cdot\vec{r}}$ factor for $|\vec{q}||\vec{r}| \gg 1 $  suppresses the growth and enforces 
the vanishing of the integral. 
Thus, we have successfully reconfirmed the $P$-condition 
from the $\vec{q}$-space point of view.

\section{Spinning 1PM} \label{sec:spin-1pm}

We generalize our construction of the boost generator to include spins of the two bodies. 
The fact that the amplitude admits a Fourier integral representation will greatly simplify the computations.

\subsection{Hamiltonian} \label{sec:spin-Hamiltonian}

\paragraph{Review of the COM Hamiltonian}
In the COM frame, the complete 1PM Hamiltonian for a spinning binary was given in \cite{Chung:2020rrz}. 
\begin{align}
    H^{[1]}|_\mathrm{COM} =-\frac{4 \pi G m_1^2 m_2^2}{E_1 E_2} \int_{\vec{q}} \frac{e^{i \vec{q} \cdot \vec{r}}}{\vec{q}^2} \left[\frac{1}{2} \sum_{s= \pm 1} e^{2 s \rho } W_1\left(s \tau_1\right) W_2\left(s \tau_2\right)\right] (U_1 U_2)_\mathrm{COM} \,.
    \label{H1-spin-COM}
\end{align}
The variable $\rho$ is the relative rapidity: $\gamma = \cosh\rho$. 
The arguments of the $W$ functions are 
\begin{align}
    \tau_a =i \frac{\varepsilon\left(q, v_1, v_2, a_a\right)}{\sinh \rho}, \quad \varepsilon(a, b, c, d)=\varepsilon_{\mu \nu \rho \sigma} a^\mu b^\nu c^\rho d^\sigma \,, 
    \quad 
    v_{a} = p_{a}/m_a \,.
    \label{vertex-tau}
\end{align}
The 4-vector $q$ is defined in the same way as in the previous section. For reasons explained there, we can set $q = (0, \vec{q})$ in the COM frame. The 4-vectors $a_a$ are the spin-length vectors of the two bodies. Their spatial components are given by $\vec{a}_a = \vec{S}_a/m_a$ with the canonical spin variables defined in \eqref{3d-Poisson}. In the COM frame the time components $(a_a)^0$ drop out completely, so their values are immaterial. Later in this subsection, we will explain how to determine $(a_a)^0$ in a general lab frame in terms of $\vec{a}_a$ and $\vec{p}_a$. 

The $W$ functions encode the general 3-point vertex with two massive legs and a single graviton, 
following the spin-multipole expansion of \cite{Levi:2015msa}:
\begin{align}
    W(\tau)=\sum_{n=0}^{\infty} \frac{C_n}{n !} \tau^n \,,
    \quad 
    C_{2k}=C_{\mathrm{ES}^{2 k}}\,,
    \quad 
    C_{2k+1}=C_{\mathrm{BS}^{2k+1}}\quad (k \geq 1) \,, 
    \quad
    C_0 = C_1 = 1\,.
\end{align}
In their original incarnation \cite{Levi:2015msa}, the ``Wilson coefficients"  $C_{2k}=C_{\mathrm{ES}^{2 k}}$ and $C_{\mathrm{BS}^{2k+1}}$ encode 
the spin-induced non-minimal couplings of a massive particle to derivatives of the curvature tensor through the worldline Lagrangian of the form (subscript SI for ``spin-induced") 
\begin{align}
\begin{split}
  &L_{\mathrm{SI}} = \sum_{n=1}^{\infty} \frac{(-1)^n}{(2 n) !} \frac{C_{ \mathrm{ES}^{2 n}}}{m^{2 n-1}} D_{\mu_{2 n}} \cdots D_{\mu_3} \frac{E_{\mu_1 \mu_2}}{\sqrt{u^2}} S^{\mu_1} S^{\mu_2} \cdots S^{\mu_{2 n-1}} S^{\mu_{2 n}} 
  \\
  &\qquad + \sum_{n=1}^{\infty} \frac{(-1)^n}{(2 n+1) !} \frac{C_{ \mathrm{BS}^{2 n+1}}}{m^{2 n}} D_{\mu_{2 n+1}} \cdots D_{\mu_3} \frac{B_{\mu_1 \mu_2}}{\sqrt{u^2}} S^{\mu_1} S^{\mu_2} \cdots S^{\mu_{2 n-1}} S^{\mu_{2 n}} S^{\mu_{2 n+1}} \,, 
  \\
  & E_{\mu \nu} \equiv R_{\mu \alpha \nu \beta} u^\alpha u^\beta \,,\quad 
  B_{\mu \nu} \equiv \frac{1}{2} \epsilon_{\alpha \beta \gamma \mu} R^{\alpha \beta}{}_{\delta \nu} u^\gamma u^\delta \,, 
  \quad 
  u^\mu = \frac{dx^\mu}{d\sigma} \,.
\end{split}
\end{align}
The $U$ functions in \eqref{H1-spin-COM} account for the Thomas-Wigner rotation factor 
in the COM frame,
\begin{align}
      (U_a)_\mathrm{COM} &= \exp \left[-i \frac{m_1 m_2}{(m_a+E_a) E}\varepsilon(q, v_1, v_2, a_a) \right] \,.
    \label{TW-COM}
\end{align}
Our goal in this section is to generalize the Hamiltonian \eqref{H1-spin-COM} to a form valid 
in an arbitrary inertial frame. The ``propagator" generalizes as $q^2 = \vec{q}^2 - (\vec{q}\cdot\vec{u}_c)^2$ 
just as in the previous section. We should discuss how the $U$ (rotation) and $W$ (vertex) functions should be generalized. 

\paragraph{Thomas-Wigner rotation in a lab frame}
The derivation of the rotation factor $U_a$ $(a=1,2)$ in \cite{Chung:2020rrz} 
is based on the relative orientation of three unit-normalized time-like vectors: 
the 4-velocity of the incoming particle $a$, that of the outgoing particle $a$, and the time axis of the reference frame common to both particles. 
The COM frame is defined such that the time axis is defined by the 4-vector
\begin{align}
  w_\mathrm{COM} =   \frac{p_1 + p_2}{\sqrt{-(p_1+p_2)^2}} \,.
\end{align}
To work in an arbitrary lab frame, we simply choose a frame and declare that the time axis is $l = (1,\vec{0})$ in that frame.


We follow \cite{Chung:2020rrz} to derive the  rotation factor in the lab frame. 
For any pair of future-pointing unit-normalized time-like vectors, it is possible to construct the minimal boost generator 
which takes one of the vectors to the other. When we have three such vectors $u$, $v$ and $w$, the chain of minimal boosts 
from $u$ to $v$ to $w$ back to $u$ leads to a net rotation in the 3d space transverse to $u$. 
As explained in \cite{Chung:2020rrz}, the magnitude of the angle can be written as a function of the three vectors. 
After accounting for 
the difference in the metric signature convention between \cite{Chung:2020rrz} and this paper, 
the formula for the rotation angle reads 
\begin{align}
    1-\cos\alpha = \frac{\left(\varepsilon_{\mu \nu \rho \sigma} u^\nu v^\rho w^\sigma\right)^2}{(1 - u \cdot v)(1 - v \cdot w)(1 - w \cdot u)} \,.
\end{align}
To be specific, let us focus on particle 1. Particle 2 can be treated similarly. 
We choose the three time-like unit vectors to be 
\begin{align}
    u = \frac{p_1 - \hbar q/2}{m_1} \,,
    \quad 
    v = \frac{p_1 + \hbar q/2}{m_1} \,,
    \quad 
    w = l = (1,\vec{0}) \,.
\end{align}
For the denominator, we have 
\begin{align}
\begin{split}
    1 - u\cdot v &= 2 + (\hbar q)^2/4m_1^2 \,,
    \\
    1 - v \cdot w &= 1 + E_1/m_1 + \hbar q^0/2m_1 \,,
    \\ 
    1 - u \cdot w &= 1 + E_1/m_1 - \hbar q^0/2m_1 \,.
\end{split}
\end{align}
In the classical limit, the $(\hbar q)$ terms are all suppressed. 
Moreover, the $q^2$ term in the first line, when expanded, will cancel against the propagator 
and produce delta-function localized terms that are irrelevant to the long-distance interaction 
between the two bodies. 

For the numerator, we have
\begin{align}
    \varepsilon_{\mu \nu \rho \sigma} u^\nu v^\rho \omega^\sigma = \pm \frac{\hbar}{m_1^2} (0,\vec{p}_1 \times \vec{q}) \,.
    \label{TW-axis}
\end{align}
Combining all the factors, and keeping only the leading term in the classical limit, we find 
\begin{align}
    2(1-\cos\alpha_1) = 4 \sin^2(\alpha_1/2) = \frac{\hbar^2 (\vec{p}_1\times\vec{q})^2}{m_1^2(E_1+m_1)^2} \,.
\end{align}
The Thomas-Wigner rotation factor is nothing but $U = \exp(i\alpha \vec{S}\cdot \hat{m}/\hbar)$, 
where $\hat{m}$ is the unit vector denoting the rotation axis given in \eqref{TW-axis}. 
Combining all elements and taking the classical ($\hbar \rightarrow 0$) limit, we obtain 
\begin{align}
    U_1 = \exp\left[ +i f_1 \vec{a}_1 \cdot (\vec{u}_1\times \vec{q}) \right] \,, 
    \quad 
    f_a = \frac{E_a}{E_a+m_a} \,, 
    \quad 
    \vec{a}_a = \frac{\vec{S}_a}{m_a} \,.
\label{TW-lab} 
\end{align}
The sign can be fixed by requiring consistency with the COM frame Hamiltonian.

\paragraph{Vertex parameter}

Let us reanalyze the $\tau$-parameters for the spin vertex in \eqref{vertex-tau}. 
Writing the Lorentz scalar in a 3d notation, we have
\begin{align}
\begin{split}
    \varepsilon\left(q, p_1, p_2, a\right) &= (E_2 \vec{p}_1 - E_1 \vec{p}_2 )\cdot (\vec{a} \times  \vec{q}) + q^0 \vec{a} \cdot (\vec{p}_1 \times \vec{p}_2) - a^0 \vec{q}\cdot (\vec{p}_1 \times \vec{p}_2) 
    \\
    &= E_1 E_2 \left[ (\vec{u}_1 - \vec{u}_2 )\cdot (\vec{a} \times  \vec{q}) + q^0 \vec{a} \cdot (\vec{u}_1 \times \vec{u}_2) - a^0 \vec{q}\cdot (\vec{u}_1 \times \vec{u}_2) \right] \,.
\end{split}
    \label{qppa-general}
\end{align}
In the COM frame, $\vec{u}_1 \times \vec{u}_2 = 0$, so the terms proportional to $q^0$ or $a^0$ cannot contribute. 
But in general, both of them are important. 
We observed in \eqref{q0-uc} that $q^0 = \vec{u}_c \cdot \vec{q}$. 
As for $a^0$, the correct assignment turns out to be 
\begin{align}
    a^0_a = \frac{\vec{p}_a\cdot\vec{S}_a}{m_a(E_a+m_a)} = f_a (\vec{u}_a \cdot \vec{a}_a) \qquad (a=1,2) \,.
    \label{a0-prescription}
\end{align}
We can understand this prescription 
by carefully studying the ``spin-gauge symmetry" 
as is reviewed in some detail in appendix~\ref{sec:spin-gauge-symm}. 
Here, we content ourselves with a short summary. 
A Lorentz covariant description of a spinning body 
begins with a spin tensor $S_{\mu\nu}$. Among its six components, only three are physical. 
There are infinitely many possible gauge choices. 
The two most popular choices are the ``covariant gauge" 
associated with the co-moving frame of the body 
and the (Pryce-Newton-Wigner \cite{Pryce:1935ibt,Pryce:1948pf,Newton:1949cq}) ``canonical gauge" associated with the lab frame.  
We define the spin-length vector $a^\mu$ such that its spatial components are $\vec{a} = \vec{S}/m$, where $\vec{S}$ is 
essentially the spatial component of the canonical spin tensor. 
Using the known map between the two gauges and the Pauli-Lubanski map from a spin tensor to a spin vector, 
we can indeed define the spin-length vector with all the desired properties, eventually leading to \eqref{a0-prescription}.

In conclusion, the all-spin 1PM Hamiltonian valid in an arbitrary inertial frame is 
\begin{align}
\begin{split}
      H^{[1]} &=-\frac{4 \pi G m_1^2 m_2^2}{E_1 E_2} \int_{\vec{q}} \frac{e^{i \vec{q} \cdot \vec{r}}}{\vec{q}^2-(q^0)^2} \left[\frac{1}{2} \sum_{s= \pm 1} e^{2 s \rho } W_1\left(s \tau_1\right) W_2\left(s \tau_2\right)\right] U_1 U_2 
    \equiv  
    \int_{\vec{q}} \widetilde{H}^{[1]} \,,
\end{split}
    \label{H1-spin-general}
\end{align}
where \eqref{TW-lab}, \eqref{qppa-general} and \eqref{a0-prescription} are all incorporated. 

\subsection{Boost} 

When we include spin, even for the free (0PM) theory, $\vec{G}$ looks non-trivial \cite{Hanson:1974qy,Bel:1980ahp}: 
\begin{align}
    H^{[0]} = E_1 + E_2 \,,
    \quad 
    \vec{G}^{[0]} = \sum_{a=1}^2 \left( E_a \vec{x}_a - \frac{\vec{S}_a \times \vec{p}_a}{E_a+m_a} \right) \,, 
    \quad E_a \equiv \sqrt{\vec{p}_a^2 + m_a^2} \,.
    \label{G0-boost}
\end{align}
At the 1PM level, we propose to work with an ansatz, 
\begin{align}
\begin{split}
     \vec{G}^{[1]}& = H^{[1]} ( \vec{X} + \vec{Y} )   \,, \quad \vec{X} = \frac{E_2 \vec{x}_1 + E_1 \vec{x}_2}{E_1+E_2}
     \\
     \vec{Y} &= \beta_1 (\vec{a}_1 \times \vec{u}_1) + \beta_2 (\vec{a}_2 \times \vec{u}_2) \,,
\end{split}
\label{G1-spin-guess}
\end{align}
where $H^{[1]}$ is the exact 1PM Hamiltonian containing all spin interactions, 
and $\beta_{1,2}$ are dimensionless functions of $E_{1,2}$ 
and $m_{1,2}$
(independent of $\vec{x}_{1,2}$ or $\vec{a}_{1,2}$).

Again, the ansatz for $\vec{Y}$ in \eqref{G1-spin-guess} is substantially simpler than the most general ansatz one can write down under the symmetry requirements. 
%
%
\textcolor{blue}{As discussed in section~\ref{sec:ansatz}, provided that our assumption on the uniqueness of $\vec{G}$ is valid, working with a simpler ansatz can be justified as long as it admits a solution.}
%
%
An attempt to gain some deeper understanding of this remarkable simplicity is beyond the scope of this paper.

We will follow the strategy outlined in the second half of section~\ref{sec:ns-1pm-G}, 
to solve the $J$-condition first and then to verify the $P$-condition in the $\vec{q}$-space. 
Following \eqref{J-1PM-new} and separating the non-spinning terms from the spinning ones, we obtain 
\begin{align}
    \{ H^{[0]} , \vec{X} \} \times \vec{Y} + \{ \vec{G}^{[0]}_\mathrm{s} , \vec{X} \}_{\times} + 
    \{ \vec{G}^{[0]}_\mathrm{n.s.} , \vec{Y} \}_{\times} + \{ \vec{G}^{[0]}_\mathrm{s} , \vec{Y} \}_{\times} = 0 \,, 
    \label{Y-eq} 
\end{align}
where the subscripts ``s" and ``n.s." stand for ``spin" and ``no spin", respectively.
Some partial computations to solve \eqref{Y-eq} include 
\begin{align}
\begin{split}
    \{ H^{[0]} , \vec{X} \} &= -z_2 \vec{u}_1 -z_1 \vec{u}_2\,,
     \\
     \{  \vec{G}^{[0]}_\mathrm{s} , \vec{X} \}_{\times} &= 
     -z_2 \left( \frac{2\vec{S}_1}{E_1 +m_1} + \frac{(\vec{S}_1 \times \vec{p}_1)\times \vec{u}_1}{(E_1 +m_1)^2}\right)
           + (1 \leftrightarrow 2) 
      \\
      &= -\frac{z_2}{E_1} \left( \vec{S}_1 + \frac{(\vec{S}_1\cdot\vec{p}_1) \vec{p}_1}{(E_1 +m_1)^2}\right)
      -\frac{z_1}{E_2} \left( \vec{S}_2 + \frac{(\vec{S}_2\cdot\vec{p}_2) \vec{p}_2}{(E_2 +m_2)^2}\right)
 \\
      &= - z_2 \left( \frac{m_1}{E_1} \vec{a}_1 + \frac{E_1 m_1 (\vec{a}_1\cdot\vec{u}_1) \vec{u}_1}{(E_1 +m_1)^2}\right)
       + (1 \leftrightarrow 2) \,,
      \\
      \{ \vec{G}^{[0]}_\mathrm{n.s.} , \vec{Y} \}_{\times} &=
      \beta_1 \left[ \frac{E_1^2+m_1^2}{E_1^2} \vec{a}_1 + (\vec{a}_1\cdot \vec{u}_1)\vec{u}_1 \right] 
      +\beta_2 \left[ \frac{E_2^2+m_2^2}{E_2^2} \vec{a}_2 + (\vec{a}_2\cdot \vec{u}_2)\vec{u}_2 \right] 
      \\
      &\quad + E_1 \left[ (\nabla_{E_1} \beta_1) \vec{u}_1 \times (\vec{a}_1\times \vec{u}_1) + (\nabla_{E_1} \beta_2) \vec{u}_1 \times (\vec{a}_2\times \vec{u}_2)\right]  
      \\
      &\quad + E_2 \left[ (\nabla_{E_2} \beta_1)  \vec{u}_2 \times (\vec{a}_1\times \vec{u}_1) + (\nabla_{E_2} \beta_2) \vec{u}_2 \times (\vec{a}_2\times \vec{u}_2)\right] \,,
      \\
      \{ \vec{G}^{[0]}_\mathrm{s} , \vec{Y} \}_{\times} &= -2\beta_1 \frac{E_1}{E_1+m_1} (\vec{a}_1\cdot \vec{u}_1)\vec{u}_1  -2\beta_2 \frac{E_2}{E_2+m_2} (\vec{a}_2\cdot \vec{u}_2)\vec{u}_2 \,.  
\end{split}
\end{align}
The cancellation of the ``cross terms" such as 
    $\vec{u}_1 \times (\vec{a}_2 \times \vec{u}_2)$
or 
    $\vec{u}_2 \times (\vec{a}_1 \times \vec{u}_1)$  
gives a set of simple differential equations 
that suggest a separation of variables of the form 
\begin{align}
    \beta_1 = z_2  h_1(E_1) \,,
    \quad 
    \beta_2 = z_1  h_2(E_2) \,.
\end{align}
Demanding that the coefficients of $\vec{a}_a$ and $(\hat{a}_a\cdot \vec{u}_a)\vec{u}_a$ $(a=1,2)$ separately vanish, we find the \emph{unique} solution to \eqref{Y-eq} given by 
\begin{align}
    h_a(E_a) = f_a = \frac{E_a}{E_a+m_a} \,. 
\end{align}
Interestingly, it is our third time to encounter the $f_a$ factors after \eqref{TW-lab} and \eqref{a0-prescription}. 

\paragraph{P-condition in q-space} 

Our Hamiltonian and boost generator take simpler form in the $\vec{q}$-space. 
%
We begin with rewriting the 0PM boost generator slightly, 
\begin{align}
   \begin{split}
\vec{G}^{[0]} &= H^{[0]} \vec{Z}^{[0]} =  H^{[0]} (\vec{X}^{[0]} + \vec{Y}^{[0]} ) \,,
       \\
       &\quad 
       \vec{X}^{[0]} = z_1 \vec{x}_1 + z_2 \vec{x}_2 \,,
       \\
       &\quad
       \vec{Y}^{[0]} = - \frac{1}{E_1+E_2} \left[ m_1 f_1 (\vec{a}_1 \times \vec{u}_1) + m_2 f_2 (\vec{a}_2 \times \vec{u}_2) \right] \,. 
    \end{split}
\end{align}
Using similar notations, 
we can write the 1PM generators, schematically, as 
\begin{align}
    \begin{split}
       \widetilde{H}^{[1]} &= \frac{k}{E_1 E_2} \frac{e^{i\vec{q}\cdot \vec{r}}}{\vec{q}^2 - (q^0)^2} W_1 W_2 U_1 U_2 = 
      \widetilde{H}^{[1]}_\mathrm{n.s.} \mathcal{S} \,,
       \\
       \vec{G}^{[1]} &=  H^{[1]} \vec{Z}^{[1]} =  H^{[1]} (\vec{X}^{[1]} + \vec{Y}^{[1]} ) \,,
       \\
       &\quad 
       \vec{X}^{[1]} = z_2 \vec{x}_1 + z_1 \vec{x}_2 \,,
       \\
       &\quad
       \vec{Y}^{[1]} = z_2 f_1 (\vec{a}_1 \times \vec{u}_1) + z_1 f_2 (\vec{a}_2 \times \vec{u}_2) \,. 
    \end{split}
\end{align}
Again, the subscript ``n.s." denotes ``no spin"; $\widetilde{H}^{[1]}_\mathrm{n.s.}$ here is the same as $\widetilde{H}^{[1]}$ in \eqref{H1-q-space}.

From the $\vec{q}$-space perspective, proving the $P$-condition amounts to 
\begin{align}
    (\vec{Z}^{[0]} - \vec{Z}^{[1]})  \{ H^{[0]}  , \widetilde{H}^{[1]} \} + H^{[0]}\{\vec{Z}^{[0]} , \widetilde{H}^{[1]} \} + \widetilde{H}^{[1]}\{\vec{Z}^{[1]} , H^{[0]} \} = \nabla_{\vec{q}} \mathcal{F} \,, 
    \label{P-HZHZ} 
\end{align}
for some function $\mathcal{F}$ to be found. 
Some parts of the LHS are easy to compute. For example, 
\begin{align}
\begin{split}
     \vec{Z}^{[0]} - \vec{Z}^{[1]} &=  (z_1-z_2) \vec{r} +  (\vec{Y}^{[0]} - \vec{Y}^{[1]}) \,,
     \\
     \vec{Y}^{[0]} - \vec{Y}^{[1]} &= - \frac{E_2+m_1}{E_1+E_2} f_1 (\vec{a}_1 \times \vec{u}_1) - \frac{E_1+m_2}{E_1+E_2} f_2 (\vec{a}_2 \times \vec{u}_2) \,.
\end{split}
\label{ZZYY}
\end{align}
Some factors do not require any new computation:
\begin{align}
\begin{split}
     \{ H^{[0]}  , \widetilde{H}^{[1]} \} &= \{ H^{[0]}  , \widetilde{H}^{[1]}_\mathrm{n.s.} \} \mathcal{S} \,,
     \\
     \{\vec{Z}^{[1]} , H^{[0]} \} &=  \{\vec{X}^{[1]} , H^{[0]} \} \,.
\end{split}
\end{align}
The most non-trivial in \eqref{P-HZHZ} is the second term: 
\begin{align}
\begin{split}
      \{\vec{Z}^{[0]} , \widetilde{H}^{[1]} \} &=  \{\vec{X}^{[0]} + \vec{Y}^{[0]},\widetilde{H}^{[1]}_\mathrm{n.s.} \mathcal{S} \} 
      \\
      &= \{\vec{X}^{[0]},\widetilde{H}^{[1]}_\mathrm{n.s.} \} \mathcal{S} +\{\vec{Y}^{[0]},\widetilde{H}^{[1]}_\mathrm{n.s.} \} \mathcal{S} 
      +  \{\vec{X}^{[0]} ,  \mathcal{S} \}\widetilde{H}^{[1]}_\mathrm{n.s.}  +  \{\vec{Y}^{[0]} ,  \mathcal{S} \}\widetilde{H}^{[1]}_\mathrm{n.s.} \,.
\end{split}
\end{align}
If we collect all terms independent of $\vec{Y}$ or $\{ \circ, \mathcal{S} \}$, 
our earlier result \eqref{P-q-total} gives
\begin{align} 
\begin{split}
    &(\vec{X}^{[0]} - \vec{X}^{[1]})  \{ H^{[0]}  ,\widetilde{H}^{[1]}_\mathrm{n.s.} \}  \mathcal{S} + H^{[0]}\{\vec{Z}^{[0]} ,\widetilde{H}^{[1]}_\mathrm{n.s.} \}  \mathcal{S}+\widetilde{H}^{[1]}_\mathrm{n.s.}\{\vec{X}^{[1]} , H^{[0]} \} \mathcal{S} 
    \\
    &\hskip 4cm = \mathcal{S} \nabla_{\vec{q}} \left[ -(\vec{u}_c\cdot \vec{q})\widetilde{H}^{[1]}_\mathrm{n.s.} \right] \,.
\end{split}
\end{align}
Now, collecting all terms involving $\vec{Y}$ or $\{ \circ, \mathcal{S} \}$ (or both), we obtain
\begin{align}
\begin{split}
      &(\vec{Y}^{[0]} - \vec{Y}^{[1]})  \{ H^{[0]}  ,\widetilde{H}^{[1]}_\mathrm{n.s.} \}  \mathcal{S}  
      + 
    H^{[0]}   \{\vec{Y}^{[0]},\widetilde{H}^{[1]}_\mathrm{n.s.} \} \mathcal{S} +  \{\vec{G}^{[0]},  \mathcal{S} \}\widetilde{H}^{[1]}_\mathrm{n.s.}  \,.
\end{split}
\label{P-cond-Y-S}
\end{align}
The $P$-condition \eqref{P-HZHZ} will hold if these terms add up to give 
\begin{align}
     - (\nabla_{\vec{q}} \, \mathcal{S} )(\vec{u}_c\cdot \vec{q})\widetilde{H}^{[1]}_\mathrm{n.s.} 
\label{total-der-S}
\end{align}
Let us check if it is indeed the case. 
First, \eqref{P-q-steps} and \eqref{ZZYY} give 
\begin{align}
\begin{split}
     &(\vec{Y}^{[0]} - \vec{Y}^{[1]})  \{ H^{[0]}  ,\widetilde{H}^{[1]}_\mathrm{n.s.} \}  
     \\
     &= i\vec{q}\cdot ( \vec{u}_1 - \vec{u}_2 )  \left[ \frac{E_2+m_1}{E_1+E_2} f_1 (\vec{a}_1 \times \vec{u}_1) + \frac{E_1+m_2}{E_1+E_2} f_2 (\vec{a}_2 \times \vec{u}_2) \right]\widetilde{H}^{[1]}_\mathrm{n.s.} \,.
\end{split}
\end{align}
We need to compute the three new contributions, 
\begin{align}
\begin{split}
      &H^{[0]}  \{\vec{Y}^{[0]},\widetilde{H}^{[1]}_\mathrm{n.s.} \}   \,,
      \\
      &H^{[0]}  \{\vec{X}^{[0]} ,  \mathcal{S} \}  =   
      \{\vec{G}^{[0]}_X ,  \mathcal{S} \}\,,
      \\
       &H^{[0]}  \{\vec{Y}^{[0]} ,  \mathcal{S} \}  =  
       \{\vec{G}^{[0]}_Y ,  \mathcal{S} \}  \,.
\end{split}
\label{new3}
\end{align}
where, in the second and third lines, we used the fact that $\mathcal{S}$ is independent of $\vec{x}_a$. 
The first item in \eqref{new3} is straightforward:
\begin{align}
\begin{split}
      H^{[0]}  \{\vec{Y}^{[0]},\widetilde{H}^{[1]}_\mathrm{n.s.} \} &= i\left[ \frac{ m_1 (\vec{a}_1\times \vec{q})}{E_1+m_1} -  \frac{m_2 (\vec{a}_2\times \vec{q})}{E_2+m_2}\right] \widetilde{H}^{[1]}_\mathrm{n.s.} 
      \\
      &\quad +   i\frac{m_1 f_1 (\vec{a}_1\times \vec{u}_1)}{E_1+E_2}\left[ (\vec{u}_2 -\vec{u_1}) \cdot\vec{q} - \left( \frac{E_1+E_2}{E_1+m_1} \right) (\vec{u}_1 \cdot\vec{q}) \right] \widetilde{H}^{[1]}_\mathrm{n.s.} 
      \\
      &\quad -  i\frac{m_2 f_2 (\vec{a}_2\times \vec{u}_2)}{E_1+E_2}\left[ (\vec{u}_1 -\vec{u_2})\cdot\vec{q} - \left(  \frac{E_1+E_2}{E_2+m_2} \right) (\vec{u}_2 \cdot\vec{q}) \right] \widetilde{H}^{[1]}_\mathrm{n.s.}\,.
\end{split}
\label{Y0-H1} 
\end{align}
When computing the second and third terms in \eqref{new3}, we should consider the amplitude factors $W_a$ and the rotation factors $U_a$ separately. 
\begin{align}
    \mathcal{S} =  \mathcal{W} \, \mathcal{U} \,,
    \quad 
     \mathcal{W}  = W_1 W_2 \,,\quad  \mathcal{U} = U_1 U_2 \,.
\end{align}

\paragraph{Rotation factor}
Let us consider $\mathcal{U}$ first. Using the explicit expression, 
\begin{align}
    \mathcal{U} = \exp\left[ i f_1 (\vec{a}_1 \times \vec{u}_1)\cdot\vec{q} - i f_2 (\vec{a}_2 \times \vec{u}_2)\cdot\vec{q} \, \right] \,,
\end{align}
we obtain some partial results, 
\begin{align}
\begin{split}
     \{ \vec{G}^{[0]}_X , \mathcal{U} \} &= i \left[ f_1 (\vec{q}\times \vec{a}_1) - f_2(\vec{q}\times \vec{a}_2)  \right] \mathcal{U} 
    \\
    &\quad  - i \left[ f_1^2 \vec{u}_1 (\vec{q}\times \vec{a}_1)\cdot \vec{u}_1 - f_2^2 \vec{u}_2 (\vec{q}\times \vec{a}_2)\cdot \vec{u}_2 \right] \mathcal{U}  \,, 
    \\
    \{ \vec{G}^{[0]}_Y , \mathcal{U} \} &= -i \left[ f_1^2 (\vec{a}_1 \cdot \vec{u}_1) (\vec{q}\times \vec{u}_1) - f_2^2 (\vec{a}_2 \cdot \vec{u}_2) (\vec{q}\times \vec{u}_2)  \right] \mathcal{U} \,.
\end{split}
\end{align}
The sum of the two exhibits some cancellation, 
\begin{align}
\begin{split}
    \{ \vec{G}^{[0]} , \mathcal{U} \} &= -i \left[ \frac{m_1(\vec{a}_1 \times \vec{q})}{E_1+m_1} - \frac{m_2(\vec{a}_2\times\vec{q})}{E_2+m_2}\right] \mathcal{U} 
   \\
   &\quad -i \left[ 
   f_1^2 (\vec{u}_1\cdot \vec{q}) (\vec{a}_1\times \vec{u}_1)-f_2^2 (\vec{u}_2\cdot \vec{q}) (\vec{a}_2\times \vec{u}_2) \right] \mathcal{U}  \,.  
\end{split}
\label{Z0-SU}
\end{align}
Between \eqref{Y0-H1} and \eqref{Z0-SU}, we observe that the $(\vec{a}_a\times \vec{q})$ terms cancel out completely. 
All the other terms are proportional to $(\vec{a}_a \times \vec{u}_a)$. Collecting the coefficients, we find 
\eqref{P-cond-Y-S} with $\mathcal{S}$ replaced by $\mathcal{U}$ 
produces precisely \eqref{total-der-S} with $\mathcal{S}$ replaced by $\mathcal{U}$.

\paragraph{Amplitude factor}

It remains to take the $\mathcal{W}$ factor into account. 
All we really need is 
\begin{align}
   \{  \vec{G}^{[0]} , \varepsilon(q,p_1,p_2,a_a) \} = \{ \vec{G}^{[0]}_X + \vec{G}^{[0]}_Y , \varepsilon(q,p_1,p_2,a_a) \}
   \quad (a=1,2) \,.
\end{align}
The computation tends to be lengthy. 
We provide some details here.
To be specific, we set $\vec{a} = \vec{a}_1$ from here on. 
The case $\vec{a} = \vec{a}_2$ can be done similarly.
We begin with breaking the ``$\tau$-parameter" into pieces:
\begin{align}
\begin{split}
      T &\equiv  \varepsilon(q,p_1,p_2,a_1) = T_A + T_B + T_C\,,
        \\
        &\qquad T_A = E_1 E_2 (\vec{u}_1 - \vec{u}_2)\cdot(\vec{a}_1\times \vec{q}) \,,
        \\
        &\qquad T_B = E_1 E_2 (\vec{u}_c \cdot \vec{q}) \vec{a}_1\cdot (\vec{u}_1 \times \vec{u}_2) \,,
        \\
        &\qquad T_C = -E_1 E_2 f_1 (\vec{u}_1\cdot \vec{a}_1) \vec{q}\cdot (\vec{u}_1 \times \vec{u}_2) \,.%
\end{split}
\end{align}
The $\vec{X}$ part of the computation shows 
\begin{align}
\begin{split}
    \{ \vec{G}^{[0]}_X , T_A \} 
   &=  E_1 E_2 \left[ (\vec{u}_1 \cdot(\vec{a}_1\times \vec{q})) \vec{u}_2 - (\vec{u}_2 \cdot(\vec{a}_1\times \vec{q})) \vec{u}_1 \right] \,,
   \\
     \{ \vec{G}^{[0]}_X, T_B \} &=  E_1 E_2\left[  (\vec{u}_c\cdot \vec{q}) \vec{a}\times (\vec{u}_1 - \vec{u}_2) 
     + (\vec{q} - \vec{u}_c (\vec{u}_c\cdot\vec{q}) ) \vec{a}_1\cdot(\vec{u}_1 \times \vec{u}_2) \right] \,,
     \\
       \{ \vec{G}^{[0]}_X, T_C \} &= -E_1E_2 f_1  (\vec{u}_1 \cdot \vec{a}_1) \left[\vec{q}\times (\vec{u}_1 - \vec{u}_2) - f_1\vec{u}_1  (\vec{q}\cdot(\vec{u}_1 \times \vec{u}_2)) \right]  
    \\
    &\qquad - E_1E_2 f_1 (\vec{q}\cdot(\vec{u}_1 \times \vec{u}_2))\vec{a}_1 \,.
\end{split}
\label{TW-X}
\end{align}
Adding the first two and using the 3d Schouten identity, we 
obtain 
\begin{align}
\begin{split}
    \{ \vec{G}^{[0]}_X, T_A+ T_B \} 
     &= (\vec{u}_c\cdot \vec{q}) E_1 E_2\left[   \vec{a}_1\times (\vec{u}_1 - \vec{u}_2) 
      - \vec{u}_c ( \vec{a}_1\cdot(\vec{u}_1 \times \vec{u}_2)) \right] 
     \\
     &\qquad + E_1 E_2 (\vec{q}\cdot(\vec{u}_1 \times \vec{u}_2))\vec{a}_1
     \\
     &= - (\vec{u}_c\cdot \vec{q}) \nabla_{\vec{q}} (T_A +T_B)  + E_1 E_2 (\vec{q}\cdot(\vec{u}_1 \times \vec{u}_2))\vec{a}_1 \,. 
\end{split}
\label{TW-Xb}
\end{align}
The $\vec{Y}$ part of the computation gives 
\begin{align}
\begin{split}
       \{ \vec{G}^{[0]}_Y , T_A \} &=  E_1 E_2 f_1 \left[ (\vec{u}_1 \cdot \vec{a}_1) \vec{q} \times (\vec{u}_1 - \vec{u}_2) - \vec{a}_1 (\vec{q}\cdot(\vec{u}_1\times \vec{u}_2))\right] \,,
      \\
        \{ \vec{G}^{[0]}_Y, T_B \} &=  E_1 E_2 f_1 (\vec{u}_c\cdot \vec{q}) (\vec{u}_1 \cdot \vec{a}_1) (\vec{u}_1 \times \vec{u}_2) \,,
       \\
        \{ \vec{G}^{[0]}_Y , T_C \} &= -E_1 E_2 f_1^2 \left[ (\vec{u}_1\cdot\vec{a}_1) \vec{u}_1 - (\vec{u}_1)^2\vec{a}_1 \right] \vec{q} \cdot (\vec{u}_1 \times \vec{u}_2) \,.
\end{split}
\label{TW-Y}
\end{align}
Once we add all terms in \eqref{TW-X} and \eqref{TW-Y}, 
while paying attention to \eqref{TW-Xb}, 
many terms cancel out. 
In particular, 
the $\vec{a}_1$ term vanishes 
through the relation $1 -2f_1 +f_1^2(\vec{u}_1)^2 = 0$. 
In the end, we confirm the desired property
\begin{align}
    \{ \vec{G}^{[0]} , T \} = - (\vec{u}_c \cdot \vec{q}) \nabla_{\vec{q}} (T) 
    \quad \Longrightarrow \quad 
     \{ \vec{G}^{[0]} , \mathcal{W} \} = - (\vec{u}_c \cdot \vec{q}) \nabla_{\vec{q}} \mathcal{W} \,.
\end{align}

\newpage 
\section{Discussion and Outlook} \label{sec:discussion}

The most obvious extension of this work is to proceed to higher orders in the PM expansion. 
In the non-spinning case, the Hamiltonian is known up to the 4PM order \cite{Bern:2021dqo,Bern:2021yeh}. 
It would be interesting to see if we can overcome the difficulties at higher orders, such as radiation reaction effects (see {\it e.g.} \cite{Kalin:2020mvi,Kalin:2020fhe,Damour:2020tta,DiVecchia:2021bdo,Bini:2021gat,Bern:2021yeh,Dlapa:2021vgp,Kalin:2022hph,Dlapa:2022lmu,Bini:2022enm,Dlapa:2023hsl} for the effects at the 3PM and 4PM orders)  
and succeed in constructing the boost generators. 
The construction of the 2PM boost generators is underway.  

In the spinning case, the all-spin classical amplitudes at the 2PM order were recently presented in \cite{Alessio:2023kgf,Aoude:2023vdk} with some apparent discrepancy between the two (see also \cite{Bautista:2023szu}) . 
By a ``classical amplitude" we mean the classical limit of a quantum amplitude 
from which a Hamiltonian can be derived. 
We expect that the boost generators will continue to serve as 
a strong consistency check and help resolve the discrepancy between \cite{Alessio:2023kgf} and \cite{Aoude:2023vdk}. 

Another obvious direction is to go from the binary dynamics to the $N$-body dynamics. A detailed discussion for $N$-body PM Hamiltonian can be found {\it e.g.} in \cite{Jones:2022aji}. 
In particular, it would be interesting to see how one of our main findings, $\vec{X}^{[1]} = z_2 \vec{x}_1 + z_1 \vec{x}_2$, should be generalized in the $N$-body problem. 

The boost generators are known to be closely related to the old subject of how to define a ``relativistic center-of-momentum coordinate". See {\it e.g.} \cite{Rothe:2010jj,Georg:2015afa,universe6020024} for recent discussions. 
It would be interesting to revisit the question with our PM generators and see how the formula for the 
``canonical center" of \cite{Rothe:2010jj,Georg:2015afa} simplifies with our exact-in-velocity formulas in isotropic coordinates. 

We conclude the discussion with a speculation on a new computational model
for the spinning PM Hamiltonian. 
There are many effective field theory models which compute the PN or PM Hamiltonian.  
It would be nice to build a new model (or slightly modify an old one) which could produce directly the all-spin Hamiltonian in the $\vec{q}$-space as in \eqref{H1-spin-general} 
and its higher PM order generalization, 
possibly without ever referring to a quantum amplitude at intermediate steps.

\acknowledgments

This work was supported in part by the National Research Foundation of Korea grant NRF-2019R1A2C2084608. 
We are grateful to Antal Jevicki, Joon-Hwi Kim, Jung-Wook Kim, Bum-Hoon Lee, Kanghoon Lee and Piljin Yi for discussions. 
We thank the Asia Pacific Center for Theoretical Physics for hospitality where a part of this work was done. 
SL is especially grateful to Michèle Levi for a series of discussions that led to the inception of this work. 
We are also grateful to two anonymous referees whose comments helped us to improve the manuscript in many aspects. 

\newpage 
\appendix

\newpage 
\section{Spin gauge symmetry} \label{sec:spin-gauge-symm}

We review the spin gauge symmetry mainly following \cite{Kim:2021rda} 
and use it to explain how the canonical spin vector $\vec{S}$ follows from the spin tensor $S_{\mu\nu}$ 
after a gauge-fixing. 

The dynamical variables of the spherical top model consists of $x^\mu$, $p_\mu$, $S_{\mu\nu}$ and 
the ``body-attached orthonormal frame", 
\begin{align}
    \eta_{\mu\nu} \Lambda^\mu{}_A  \Lambda^\nu{}_B = \eta_{AB} \,,\qquad  \Lambda^\mu{}_A  \Lambda^\nu{}_B \eta^{AB} = \eta^{\mu\nu} \,.
    \label{BAOF}
\end{align}
The non-vanishing Poisson brackets are 
\begin{align}
\begin{split}
\{ x^\mu , p_\nu \} &= \delta^\mu{}_\nu \,, 
\\
\{  \Lambda^\rho{}_A, S_{\mu\nu} \} &= -( \delta^\rho{}_\mu  \Lambda_{\nu A} - \delta^\rho{}_\nu \Lambda_{\mu A} )
\,,
\\
\{ S_{\mu\nu} , S_{\rho\sigma} \} &= 
- (\eta_{\nu\rho} S_{\mu \sigma} - \eta_{\mu\rho} S_{\nu\sigma} 
- \eta_{\nu\sigma} S_{\mu\rho} + \eta_{\mu\sigma}S_{\nu\rho} ) 
\,.
\label{poisson-1}
\end{split}
\end{align}
The mass-shell constraint and the spin-gauge generators are 
\begin{align}
    \phi_0 = \frac{1}{2}(p^2 + m^2) \,,
    \quad 
    \phi_a = \frac{1}{2}(\hat{p}^\mu + \Lambda^\mu{}_0) S_{\mu\nu} \Lambda^\nu{}_a \,, 
    \quad 
    \hat{p}^\mu = p^\mu/|p| \,.
\end{align}

\paragraph{Gauge-fixing constraints}

Ref.~\cite{Kim:2021rda} studied the covariant gauge-fixing constraints, 
\begin{align}
    (\chi_\mathrm{cov})^0=\frac{1}{p^2} x^\mu p_\mu, \quad (\chi_\mathrm{cov})^a=\hat{p}_\mu \Lambda^{\mu a} \,.
    \label{gf-cov}
\end{align}
These constraints \emph{imply} the covariant spin condition:
\begin{align}
    \Lambda^\mu{}_0 = \hat{p}^\mu \,,
    \quad 
    S_{\mu\nu}p^\nu = 0\,.
\end{align}

To analyze a binary system, it is often convenient to work in a fixed lab frame, 
which will lead us to the ``canonical" Pryce-Newton-Wigner gauge \cite{Pryce:1935ibt,Pryce:1948pf,Newton:1949cq}. 
Given the lab frame 4-vector, $l = (1,\vec{0})$, we may consider 
\begin{align}
    \chi^0 = -\frac{x \cdot l}{m}  \,,
    \quad 
    \chi^a = l_\mu \Lambda^{\mu a} \,.
    \label{gf-lab}
\end{align}
We normalized the new constraints such that 
they become equal to the covariant ones when the lab frame coincide with the co-moving frame. 
The new constraints imply that 
\begin{align}
    \Lambda^\mu{}_0 = l^\mu \,,
    \quad 
    (p^\mu + m l^\mu) S_{\mu\nu}\Lambda^{\nu}{}_a = 0\,.
    \label{lab-constraint}
\end{align}
In the lab frame, the second equation leads to 
\begin{align}
    (E + m) S_{0i} + p^j S_{ji} = 0 
    \quad 
    \mbox{or}
    \quad 
    S_{0i} = \frac{S_{ij}p^j}{E+m} = - \frac{(\vec{S}\times \vec{p})_i}{E+m} \,,
    \quad 
    S_i \equiv \frac{1}{2} \epsilon_{ijk} S^{jk} \,.
    \label{S0i-how}
\end{align}
We will shortly see how this explains the spin-part of $\vec{G}^{[0]}$.

\newpage 
\paragraph{Dirac brackets in the canonical gauge} 

Next, we compute the Poisson brackets between $\chi^A$ and $\phi_B$ and use them to compute the Dirac brackets. 
It is easy to check that 
\begin{align}
    \{\chi^A , \chi^B \} = 0 = \{ \phi_A ,\phi_B \} \,.
\end{align}
Computing $C^A{}_B \equiv \{ \chi^A , \phi_B \}$ 
up to terms that vanish under the constraints, we find 
\begin{align}
\begin{split}
     \{ \chi^0 , \phi_0 \} = \frac{E}{m} \,, &\quad \{ \chi^0 , \phi_b \} = \frac{E-m}{2m^3} (l^\mu S_{\mu\nu} \Lambda^\nu{}_b) \,, 
     \\
      \{ \chi^a, \phi_0 \} = 0 \,, &\quad  \{ \chi^a , \phi_b \} =  \frac{E+m}{2m}  \delta^a{}_b \,.
\end{split}
\label{chi-phi-PB}
\end{align}
The Dirac bracket is defined in the standard manner, 
\begin{align}
\begin{split}
      \{f, g\}_* &= \{f, g\}- (C^{-1})^A{}_B \left[ \left\{f, \phi_A\right\}\left\{\chi^B, g\right\} - \left\{f, \chi^B\right\}\left\{\phi_A, g\right\} \right] \,.
\end{split}
\end{align}
Computing the Dirac brackets 
and using the constraints to simplify the results at the end, 
\begin{align}
    \begin{split}
        \{ x^\mu , x^\nu \}_* &= 0 \,,
        \\
         \{ x^\mu , p_\nu \}_* &= \delta^\mu{}_\nu + \frac{1}{E} p^\mu l_\nu \,,
         \\
         \{ x^\rho , S_{\mu\nu} \}_* 
        &= -\frac{1}{E(E+m)} \left( p^\rho l^\sigma + E \eta^{\rho\sigma} \right) (S_{\sigma\mu}l_{\nu} - S_{\sigma\nu}l_{\mu}) \,, 
         \\
          \{ S_{\mu\nu} , S_{\rho\sigma} \}_* 
          &= \{ S_{\mu\nu} , S_{\rho\sigma} \} - \frac{4}{E+m} \left(  p_{[\mu} S_{\nu] [\rho} l_{\sigma]} 
          - p_{[\rho} S_{\sigma] [\mu} l_{\nu]}
          \right)  \,.
    \end{split}
\end{align}

We are ready to show that the Dirac bracket here agrees 
with the Poisson bracket in the main text, 
with $(\vec{x}, \vec{p}, \vec{S})$ as independent variables. 
Let us first look at the $(x,p)$ sector in the lab frame 
$l^\mu = (1,\vec{0})$: 
\begin{align}
\begin{split}
     &\{ x^0 , p_0 \}_* 
     = 0 \,,
    \qquad 
    \{ x^0 , p_i \}_* = 0 \,,
    \\
    &\{ x^i, p_0 \}_* = - \frac{p_i}{E} \,,
    \quad 
     \{ x^i, p_j \}_* = \delta^i{}_j \,. 
\end{split}
\end{align}
Next, we look at the $(x,S)$ sector.
\begin{align}
\begin{split}
    \{ x^0 , S_{\mu\nu} \}_* &= 0 \,,
    \quad 
    \{ x^\rho , S_{ij} \}_* = 0 \,.
\\
    \{ x_i  , S_{0j} \}_* &= -\frac{1}{E+m}\left(\frac{1}{E} p_i S_{jk} p_k + S_{ij} \right) = \left\{ x_i , \frac{S_{jk}p_k}{E+m} \right\}_* \,. 
\end{split}
\end{align}
Finally, the $(S,S)$ sector. The $\{S_{ij}, S_{kl}\}$ brackets remain unchanged. The other ones are 
\begin{align}
    \begin{split}
        \{ S_{ij}, S_{0k} \}_* &= \frac{1}{E+m}\left( \delta_{ik}S_{jl}p^l - \delta_{jk}S_{il}p^l -p_i S_{jk} - p_j S_{ik} \right)
        = \left\{S_{ij} , \frac{S_{kl}p^l}{E+m} \right\}_* \,,
        \\
        \{ S_{0i}, S_{0j} \}_* &= \frac{E-m}{E+m} S_{ij} + \frac{p_i S_{0j} - p_j S_{0i}}{E+m} 
        =  \left\{\frac{S_{ik}p^k}{E+m}, \frac{S_{jl}p^l}{E+m} \right\}_* \,.
    \end{split}
\end{align}

\paragraph{Covariant vs Canonical} 
In general, the spin-gauge transformation generates shifts 
\begin{align}
  (x')^\mu = x^{\mu} + \Delta x^\mu \,,
  \quad 
  (S')_{\mu\nu} = S_{\mu\nu} + \Delta S_{\mu\nu} \,,
\end{align}
such that both $p_\mu$ and 
$J_{\mu\nu} = x_\mu p_\nu - x_\nu p_\mu + S_{\mu\nu}$
remain invariant. In other words, 
\begin{align}
  \Delta S_{\mu\nu} + (\Delta x_\mu) p_\nu - (\Delta x_\nu) p_\mu = 0 \,. 
\end{align}

In this paper, we need both of the two famous gauge choices: the covariant one for the co-moving frame and the canonical one for the lab frame.  
To avoid confusion, let us denote the covariant spin tensor by $\Sigma_{\mu\nu}$. Consider the spatial parts:
\begin{align}
    \Sigma_k \equiv \frac{1}{2} \epsilon_{ijk} \Sigma_{jk} \,,
    \quad 
    S_k = \frac{1}{2} \epsilon_{ijk} S_{jk} \,. 
\end{align}
The time components ($S_{0i}$ and $\Sigma_{0i}$) are determined by the space components ($S_{ij}$ and $\Sigma_{ij}$) through the respective gauge conditions. Explicitly, 
\begin{align}
    \Sigma_{0i} = -\frac{(\vec{\Sigma}\times \vec{p})_i}{E}\,,
    \quad 
    S_{0i} = -\frac{(\vec{S}\times \vec{p})_i}{E+m} \,.
\end{align}

Eq.(3.60b) of \cite{Hanson:1974qy} gave the relation between the two, which was reproduced in the spin-gauge language in eq.(2.16) of \cite{Kim:2021rda}:
\begin{align}
        \vec{\Sigma} = \frac{E}{m} \vec{S} - \frac{\vec{p}(\vec{p}\cdot\vec{S})}{m(E+m)} 
\quad 
\Longleftrightarrow
\quad 
    \vec{S} = \frac{m}{E} \vec{\Sigma} + \frac{\vec{p}(\vec{p}\cdot\vec{\Sigma})}{E(E+m)} \,.
        \label{S-vs-Sigma}
\end{align}
It is useful to note that
\begin{align}
    \vec{p} \cdot \vec{\Sigma} =  \vec{p}\cdot \vec{S} \,,
    \quad 
    \frac{\vec{p} \times \vec{\Sigma}}{E} = \frac{\vec{p} \times \vec{S}}{m} \equiv \vec{p} \times \vec{a} \,, 
    \quad 
    \frac{\vec{\Sigma}}{E} = \vec{a} - f \vec{u}(\vec{u}\cdot\vec{a}) \,, 
    \quad 
    f= \frac{E}{E+m} \,.
    \label{pspsps}
\end{align}
Another important fact explained in \cite{Hanson:1974qy} is that 
\begin{align}
    \vec{S}^2 = \frac{1}{2} \Sigma^{\mu\nu} \Sigma_{\mu\nu} \,.
\end{align}
The RHS is both Lorentz invariant and spin-gauge invariant, 
while the LHS appears to be Lorentz non-invariant. 
The spin-gauge symmetry accounts for the apparent mismatch.  

We are ready to explain the origin of the ``$a^0$ prescription" used in the main text. 
The original spin variables ($S_{\mu\nu}$ or $\Sigma_{\mu\nu}$) 
are tensors while $a^\mu$ is a vector. 
The standard procedure to switch between a spin tensor 
and a spin vector is the Pauli-Lubanski map, which 
is invertible only in the covariant gauge. In our convention, the map is 
\begin{align}
    W_\mu = - \frac{1}{2m} \varepsilon_{\mu\nu\rho\sigma} \Sigma^{\nu\rho}p^\sigma 
    \quad 
    \Longleftrightarrow 
    \quad 
    \Sigma_{\mu\nu} = \frac{1}{m} \varepsilon_{\mu\nu\rho\sigma} p^\rho W^\sigma \,.
    \label{Pauli-Lubanski}
\end{align}
Since we prefer $S$ to $\Sigma$ in the main text, we use \eqref{pspsps} to relate $W^\mu$ to $\vec{S}$ to obtain  
\begin{align}
 W^0 = \frac{\vec{p}\cdot\vec{S}}{m}\,,
 \quad 
    \vec{W} &= \vec{S} + \frac{\vec{p}(\vec{p}\cdot \vec{S})}{m(E+m)} \,.
\end{align}
We could define $a^\mu$ as $W^\mu/m$, but using the fact that 
$\varepsilon(q,p_1,p_2,a_1)  =  \varepsilon(q,p_1,p_2,a_1+\kappa p_1)$ 
for any $\kappa$, we modify the definition of $a^\mu$ such that
\begin{align}
    a^\mu \equiv \frac{W^\mu}{m} - \frac{(\vec{p}\cdot\vec{S})p^\mu }{m^2(E+m)} 
    \quad 
    \Longrightarrow
    \quad 
    \vec{a} = \frac{\vec{S}}{m} \,,
    \quad 
    a^0 = \frac{\vec{p}\cdot\vec{S}}{m(E+m)} \,.
\end{align}

\section{PN expansion of PM Hamiltonian} \label{sec:more-PN}

\paragraph{Spin-orbit coupling}

In \eqref{H1-spin-general}, we presented the general spinning 1PM Hamiltonian as a Fourier integral. 
Performing the Fourier integral to write down the Hamiltonian in the position space is conceptually straightforward but practically cumbersome. 
We illustrate the idea by presenting the explicit form of the spin-orbit Hamiltonian, linear in spins but still exact in velocities.

The spin-orbit coupling consists of two groups of terms. One comes from the generating function and is linear in $\gamma$. The other one comes from the Thomas-Wigner rotation factor and is proportional to $(2\gamma^2-1)$.
Each group is split into two pieces based on their dependence 
on the velocity vectors. 
All in all, we have
\begin{align}
\begin{split}
    H^{[1]}_{\mathrm{SO}} &= H^{[1a]}_{\mathrm{SO}} + H^{[1b]}_{\mathrm{SO}} + H^{[1c]}_{\mathrm{SO}} +
    H^{[1d]}_{\mathrm{SO}}\,,
    \\
    H^{[1a]}_{\mathrm{SO}} &=  -2\frac{G m_1 m_2}{r^2}  \gamma \gamma_c^3(1- \vec{u}_c^2)  (\vec{u}_1 - \vec{u}_2) \cdot \left[ \hat {n} \times (\vec{a}_1 + \vec{a}_2) \right] \,,
    \\
    H^{[1b]}_{\mathrm{SO}} &=  - 2\frac{Gm_1m_2}{r^2} \gamma \gamma_c^3  (1- \vec{u}_c^2)  ( \hat{n} \cdot (\vec{u}_1 \times \vec{u}_2)) \left[ f_1 (\vec{u}_1 \cdot \vec{a}_1) + f_2  (\vec{u}_2\cdot \vec{a}_2) \right]\,,
      \\
    H^{[1c]}_{\mathrm{SO}} &=  - \frac{ G m_1^2 m_2^2(2\gamma^2-1)}{E_1 E_2 r^2 } \gamma_c^3   (1- \vec{u}_c^2)  \left[ f_1 (\vec{u}_1 \times \vec{a}_1) -f_2 (\vec{u}_2 \times \vec{a}_2)\right] \cdot \hat{n} \,,
    \\
    H^{[1d]}_{\mathrm{SO}} &=  - \frac{ G m_1^2 m_2^2(2\gamma^2-1)}{E_1 E_2 r^2 } \gamma_c^3  (\vec{u}_c \cdot \hat{n}) \left[ f_1 (\vec{u}_1 \times \vec{a}_1) -f_2 (\vec{u}_2 \times \vec{a}_2)\right] \cdot \vec{u}_c \,.
\end{split}
\label{H1-SO-all}
\end{align}
Here, the $H^{[1a]}_{\mathrm{SO}}$ term comes from the first two terms of \eqref{qppa-general}. Its simplicity is a result of partial cancellations. The $H^{[1b]}_{\mathrm{SO}}$ term comes from the last term of \eqref{qppa-general}. It is simple because $\vec{u}_c\times (\vec{u}_1\times \vec{u}_2) = 0$. The $H^{[1c]}_{\mathrm{SO}}$ and $H^{[1d]}_{\mathrm{SO}}$ terms 
come from the rotation factor \eqref{TW-lab}.

Here are some useful relations for computing the SO Hamiltonian:
\begin{align}
    \nabla \left( \frac{\gamma_c}{r} \right) = - \frac{\gamma_c^3}{r^2}  \left[ (1 - \vec{u}_c^2 )\hat{n}  + (\vec{u}_c\cdot\hat{n}) \vec{u}_c \right] \,, 
    \quad 
    \vec{u}_c \times (\vec{u}_1 - \vec{u}_2) = -( \vec{u}_1 \times \vec{u}_2 )\,.
\end{align}

\paragraph{PN expansion} 

We give an explicit expression for the 2PN expansion of the SO Hamiltonian given in \eqref{H1-SO-all}.
The $[1b]$ part is the easiest. 
It is already cubic in velocities, 
so we simply make the replacements, 
\begin{align}
    \gamma \rightarrow 1\,,
    \quad 
    \gamma_c \rightarrow 1\,,
    \quad 
    1- \gamma_c^2(\vec{u}_c \cdot \hat{n})^2 \rightarrow 1 \,,
    \quad 
    \vec{u}_{1,2} \rightarrow \vec{v}_{1,2} \,,
    \quad 
    f_{1,2} \rightarrow \frac{1}{2} \,,
\end{align}
and run the 3d Schouten identity to find 
\begin{align}
\begin{split}
     H^{[1b](3)}_{\mathrm{SO}} &= -\frac{Gm_1m_2}{r^2}  \hat{n} \cdot (\vec{v}_1 \times \vec{v}_2) \left( \vec{v}_1 \cdot \vec{a}_1 + \vec{v}_2\cdot \vec{a}_2 \right)
     \\
     &= -\frac{Gm_1m_2}{r^2}\left[ (a_1,v_1,v_2)(\hat{n}\cdot \vec{v}_1) + (v_1,a_1, n)(\vec{v}_1\cdot \vec{v}_2) - (v_2,a_1, n) \vec{v}_1^2 \right]
     \\
     &\quad +\frac{Gm_1m_2}{r^2}\left[ (a_2,v_2,v_1)(\hat{n}\cdot \vec{v}_2) + (v_2,a_2, n)(\vec{v}_1\cdot \vec{v}_2) - (v_1,a_2, n) \vec{v}_2^2 \right]\,. 
\end{split}
\end{align}
The $[1a]$ part requires a bit more work:
\begin{align}
\begin{split}
    H^{[1a](3)}_{\mathrm{SO}} &= - \frac{Gm_1m_2}{r^2} \left[(\vec{v}_c^2-3(\hat{n} \cdot\vec{v}_c)^2)(\vec{v}_1-\vec{v}_2) \right. 
    \\
    &\hskip 2cm
    \left. +(\vec{v}_1-\vec{v}_2)^2 (\vec{v}_1-\vec{v}_2) -  (\vec{v}_1^2\vec{v}_1-\vec{v}_2^2\vec{v}_2)\right] \cdot [\hat{n} \times (\vec{a}_1+\vec{a}_2)]
    \\
    &= \frac{Gm_1m_2}{r^2}\left[ (\vec{v}_c^2-3(\hat{n} \cdot\vec{v}_c)^2)(\vec{v}_1-\vec{v}_2)  \right.
    \\
    &\hskip 2cm
    \left. + (\vec{v}_2^2 -2 \vec{v}_1\cdot \vec{v}_2)\vec{v}_1  -(\vec{v}_1^2 - 2 \vec{v}_1\cdot \vec{v}_2) \vec{v}_2 \right] \cdot [ (\vec{a}_1+\vec{a}_2)\times \hat{n}]
    \,.
\end{split}
\end{align}
The $[1c]$ part is the most complicated:
\begin{align}
\begin{split}
    H^{[1c](3)}_{\mathrm{SO}} &=  -\frac{Gm_1m_2}{8r^2} \left[-\vec{v}_1^2 (\vec{v}_1 \times \vec{a}_1)\cdot \hat{n}  +\vec{v}_2^2  (\vec{v}_2 \times \vec{a}_2) \cdot\hat{n} \right.
    \\
    &\hskip 2.5cm +2 (\vec{v}_c^2 -3(\hat{n} \cdot \vec{v}_c)^2 )(\vec{v}_1 \times \vec{a}_1-\vec{v}_2 \times \vec{a}_2) \cdot \hat{n}
    \\
    &\hskip 2.5cm  \left. +2(3\vec{v}_1^2+3\vec{v}_2^2-8(\vec{v}_1 \cdot \vec{v}_2) ) (\vec{v}_1 \times \vec{a}_1 - \vec{v}_2 \times \vec{a}_2)  \cdot \hat{n} \right] \,.
\end{split}
\end{align}
The $[1d]$ part is just as easy as the $[1b]$ part:
\begin{align}
\begin{split}
    H^{[1d](3)}_{\mathrm{SO}} &=  -\frac{Gm_1m_2}{2r^2} 
    (\hat{n} \cdot \vec{v}_c)(\vec{v}_1 \times \vec{a}_1-\vec{v}_2 \times \vec{a}_2) \cdot \vec{v}_c
    \\
    &= \frac{Gm_1m_2}{2r^2} 
    (\hat{n} \cdot \vec{v}_c)(\zeta_2 \vec{a}_1 + \zeta_1 \vec{a}_2) \cdot (\vec{v}_1 \times \vec{v}_2) 
      \,.
\end{split}
\end{align}
Adding up all four groups, we finally obtain 
\begin{align}
\begin{split}
    H^{[1](3)}_\mathrm{SO} &= \frac{Gm_1m_2}{8r^2} \sum_{a=1,2} A^{(3)}_a  \,,
    \\
    A^{(3)}_1 &= (a_1,v_1,n) \left[5\vec{v}_1^2 +8\vec{v}_1\cdot\vec{v}_2 -2\vec{v}_2^2 -6\vec{v}_c^2 +18(\hat{n}\cdot\vec{v}_c)^2 \right] 
    \\
    &\quad +(a_1,v_2 ,n) \left[ -16\vec{v}_1 \cdot \vec{v}_2 +8 \vec{v}_c^2 -24(\hat{n}\cdot \vec{v}_c)^2 \right]  
    \\
    &\qquad + (a_1,v_1,v_2) \left[ (4\zeta_1\zeta_2 - 8) (\hat{n}\cdot\vec{v}_1) + 4 \zeta_2^2(\hat{n}\cdot\vec{v}_2) \right] \,.
\end{split}
\end{align}
The expression for $A^{(3)}_2$ is obtained by the exchange symmetry.

\newpage
\section{Comparison with ADM} \label{sec:PN-check}

As a further consistency check for the new results we obtained in the main text, 
we show that our Hamiltonian and boost generator agree, up to a canonical transformation, with the 1PM part of the PN computations in the literature. 
Specifically, we compare our results with the ADM results up to the 2PN order.

To compare two different sets of canonical coordinates, we need to find the canonical transformation between the two. 
A continuous canonical transformation can be represented in a Lie-series form, 
generated by a smooth function $C$ on the phase space, 
\begin{align}
    T_C A=e^{\{ \circ , C\}} A \equiv A+\{A, C\}+\frac{1}{2}\{\{A, C\}, C\}+\frac{1}{6}\{\{\{A, C\}, C\}, C\} + \cdots \,
\end{align}
%
We are interested in a situation where the observable $A$ and the generator $C$ both admit (PN or PM) perturbative expansions. 
Then we can match terms of the same perturbative order. For example, in a PN comparison below we will encounter an equation such as 
\begin{align}
    \Delta H^{(2)} \equiv H^{(2)}_\mathrm{PM} - H^{(2)}_\mathrm{ADM} = \{ H^{(1)} , C^{(1)} \}  \,, 
    \quad 
    H^{(1)} = H^{(1)}_\mathrm{PM} = H^{(1)}_\mathrm{ADM} \,.
    \label{cano-perturbative} 
\end{align}
Here, PM refers to the isotropic coordinates used in this paper and ADM refers to the coordinates adopted in the ADM formulation. 
The expression $H^{(2)}_\mathrm{PM} - H^{(2)}_\mathrm{ADM}$ may look like a difference between two functions of the same variables. 
But, it originates from a single function seen from two coordinate systems (A for ADM, P for PM):
\begin{align}
    H_\mathrm{P}(x_\mathrm{P},p_\mathrm{P})  = H_\mathrm{A}(x_\mathrm{A}, p_\mathrm{A}) \,.
    \label{PPAA}
\end{align}
To avoid clutter, let us remove subscripts from $(x_\mathrm{P}, p_\mathrm{P})$ and simply write $(x,p)$. 
To the linear order in perturbation, the coordinate transformation is given by
\begin{align}
    (x_\mathrm{A},p_\mathrm{A}) = (x + \Delta x, p + \Delta p) \approx (x + \{ x, C^{(1)} \} , p + \{p, C^{(1)} \}) \,.
\end{align}
Then the RHS of \eqref{PPAA} gives 
\begin{align}
    H_\mathrm{A}(x_\mathrm{A},p_\mathrm{A}) =  H_\mathrm{A}(x + \Delta x, p + \Delta p) \approx H_\mathrm{A}(x,p) + \{ H_\mathrm{A} , C^{(1)} \} \,.
\end{align}
Moving $H_\mathrm{A}(x,p)$ to the LHS of \eqref{PPAA}, we recover \eqref{cano-perturbative}. 

\paragraph{PN order counting} 
We assign weight $(+1/2)$ to $\vec{p}_{1,2}$ and Newton's constant $G$ while assigning $(-1/2)$ to $\vec{x}_{1,2}$.  
The masses $m_{1,2}$ do not carry a PN weight. 
In this convention, the exact translation and rotation generators are denoted by 
\begin{align}
    \vec{P} = \vec{P}^{(1/2)} = \vec{p}_1 + \vec{p}_2 \,,
    \quad 
    \vec{J} = \vec{J}^{(0)} = \vec{x}_1\times \vec{p}_1 + 
    \vec{x}_2\times \vec{p}_2  \,.
\end{align}
As for the Hamiltonian, 
the Newtonian part carries weight $(+1)$ 
and the $k$-PN Hamiltonian weight carries weight $(k+1)$. 
The ``pre-Newtonian" and the Newtonian terms are 
\begin{align}
        H^{(0)} = m_1 + m_2 \,,
        \quad 
        H^{(1)} = \frac{\vec{p}_1^2}{2m_1} + \frac{\vec{p}_2^2}{2m_2} - \frac{Gm_1 m_2}{r} \,.
\end{align}
We will proceed to the 2PN order, or $H^{(3)}$, in this section. 

\subsection{Non-spinning 1PN}

Here are some short-hand notations to be used in this section:
\begin{align}
\begin{split}
    \vec{v}_a = \frac{\vec{p}_a}{m_a} \,,
    \quad 
    M = m_1 + m_2 \,,
    \quad 
    \zeta_a = \frac{m_a}{m_1+m_2} \,, 
    \quad 
    \vec{v}_c = \frac{\vec{p}_1 + \vec{p}_2}{m_1+m_2} \,.
\end{split}
\label{PN-notations}
\end{align}

\paragraph{1PN ADM Hamiltonian}

The 1PN Hamiltonian in the ADM gauge is given by
\begin{align}
\begin{split}
        H^{(2)} &= H^{(2a)} + H^{(2b)} + H^{(2c)} \,, 
        \\
        &H^{(2a)} = -\frac{\vec{p}_1^4}{8 m_1^3}-\frac{\vec{p}_2^4}{8 m_2^3} \,,
        \\
        &H^{(2b)} = \frac{Gm_1 m_2 }{2  r}\left[-3 (  \vec{v}_1^2 +  \vec{v}_2^2 ) +7  (\vec{v}_1 \cdot \vec{v}_2) +  (\hat{n} \cdot \vec{v}_1) (\hat{n} \cdot \vec{v}_2) \right]\,,
        \\
        &H^{(2c)} =  \frac{G^2m_1 m_2 }{2 r^2}\left(m_1+m_2\right) \,.
\end{split}
\label{H-PN012}
\end{align}
The leading term of the boost generator
\begin{align}
        \vec{G}^{(-1/2)} &= m_1 \vec{x}_1 + m_2 \vec{x}_2 \,,
\end{align}
satisfies all of the desired properties, 
\begin{align}
    \{ G^{(-1/2)}_i , P^{(1/2)}_j \} = \delta_{ij} H^{(0)} \,, 
    \quad
    \{ G^{(-1/2)}_i , H^{(0)} \} = 0 \,, 
    \quad
    \{ G^{(-1/2)}_i , H^{(1)} \} = P^{(1/2)}_i \,.
\end{align}

\paragraph{1PN ADM boost} 
The next order boost generator, $\vec{G}^{(1/2)}$, 
is known to be \cite{Damour:2000kk}
\begin{align}
        \vec{G}^{(1/2)} &= \frac{\vec{p}_1^2}{2m_1} \vec{x}_1 + \frac{\vec{p}_2^2}{2m_2} \vec{x}_2 + \frac{1}{2} (\vec{x}_1 + \vec{x}_2) \left(-\frac{Gm_1m_2}{r} \right) \,.
        \label{G12-Y}
\end{align}
It is straightforward to verify the $H$/$P$/$J$-conditions 
at this order. 
\begin{align}
\begin{split}
      \{ G^{(1/2)}_i , P^{(1/2)}_j \} &= \delta_{ij} H^{(1)}
   \\
    \{ \vec{G}^{(1/2)} , H^{(1)} \} + \{ \vec{G}^{(-1/2)} ,  H^{(2)} \} &= 0 \,, 
    \\
    \{ G^{(-1/2)}_i  , G^{(1/2)}_j \} + \{ G^{(1/2)}_i  , G^{(-1/2)}_j \} &= - \epsilon_{ijk} J_k \,.
    \label{G12-HPJ} 
\end{split}
\end{align}

\paragraph{1PM to 1PN}

We wish to compare our 1PM generators with the ADM results at the 1PN order. 
Expanding the Lorentz factor in the PN notation, we find 
\begin{align}
       \gamma_c &=  1 + \frac{1}{2}(\vec{v}_c)^2_\perp 
      + \frac{1}{8}(\vec{v}_c)^2_\perp \left[ 3(\vec{v}_c)^2_\perp - 4  (\zeta _1\vec{v}_1^2 + \zeta _2\vec{v}_2^2) \right] + \mathcal{O}(v^6) \,. 
\label{gammac-PN}
\end{align}
We can use it to compute the PN expansion of 
the 1PM Hamiltonian $H^{[1]}$. 
At 1PN, 
\begin{align}
\begin{split}
   H^{[1](2)} &= H^{[1](2)}_\mathrm{bare} +  \frac{1}{2} (\vec{v}_c)^2_\perp H^{[1](1)}_\mathrm{bare}  
   \\
   &= \frac{Gm_1m_2}{2r}\left[ -3 (  \vec{v}_1^2 +  \vec{v}_2^2 ) +8  (\vec{v}_1 \cdot \vec{v}_2)  -  (\vec{v}_c)^2_\perp \right] \,.
\end{split}
\label{H-1PM-1PN-full}
\end{align}
Our 1PM boost generator at the 1PN order gives 
\begin{align}
      \vec{G}^{[1](1/2)} &=  \left(-\frac{Gm_1m_2}{r} \right) (\zeta_2 \vec{x}_1 + \zeta_1 \vec{x}_2)  \,.
\label{G-1PM-1PN}
\end{align}

\paragraph{Canonical transformation from ADM to PM} 

The equivalence between the two gauges, which we call ADM and PM, can be established by 
finding a canonical transformation between the two. 
The 1PN Hamiltonian transforms between two gauges as
\begin{align}
   \Delta H^{(2)} = \{ H^{(1)}, C^{(1)} \} \,.
\end{align}
The general form of $C^{(1)}$ is 
\begin{align}
    C^{(1)} = G m_1 m_2 ( c_0 \vec{v}_1 -\bar{c}_0 \vec{v}_2 )\cdot \hat{n} \,.
    \label{C1-general} 
\end{align}
The coefficients $c_0$, $\bar{c}_0$ are dimensionless functions of mass ratios 
and mapped to each other by the exchange symmetry $m_1\leftrightarrow m_2$. 
The resulting shift of the 1PN Hamiltonian is 
\begin{align}
\begin{split}
        \Delta H^{(2)} &= 
        - \frac{Gm_1m_2}{r^2} \left[ ( c_0 \vec{v}_1 -\bar{c}_0 \vec{v}_2 )\cdot (\vec{v}_1-\vec{v}_2) \right]_\perp + \frac{G^2m_1^2m_2^2}{r^2} \left( \frac{c_0}{m_1} + \frac{\bar{c}_0}{m_2} \right) \,, 
\end{split}
\end{align}
 where we introduced a notation for the transverse part of the inner product, 
\begin{align}
    (\vec{A}\cdot\vec{B})_\perp &\equiv \vec{A}\cdot\vec{B} - (\vec{A}\cdot\hat{n})(\hat{n}\cdot\vec{B}) \,.
\end{align}
To reach the PM gauge, we choose the coefficients in \eqref{C1-general} to be 
\begin{align}
c_0 =  \frac{1}{2} \zeta_1^2 \,,
\quad 
\bar{c}_0 = \frac{1}{2} \zeta_2^2
\quad 
\Longrightarrow
\quad 
C^{(1)} = \frac{1}{2} Gm_1m_2 (\zeta_1^2 \vec{v}_1 - \zeta_2^2 \vec{v}_2)  \cdot \hat{n} \,,
\label{c1-ADM-PM}
\end{align} 
which produces, with $\vec{v}_c$ defined as in \eqref{PN-notations}, 
\begin{align}
    \Delta H^{(2b)} &=  \frac{Gm_1m_2 }{2r} \left[ (\vec{v}_1 \cdot \vec{v}_2)_\perp - (\vec{v}_c)^2_\perp \right] \,, \quad 
    \Delta H^{(2c)} = \frac{G^2 m_1^2 m_2^2}{2M^2 r^2} \,.
\label{H-1PN-cano-PM}
\end{align}
Adding this to the $\mathcal{O}(G)$ part of the ADM Hamiltonian, we obtain  
\begin{align}
    H^{(2b)} + \Delta H^{(2b)} = \frac{Gm_1 m_2 }{2  r}\left[-3 (  \vec{v}_1^2 +  \vec{v}_2^2 ) + 8  (\vec{v}_1 \cdot \vec{v}_2) - (\vec{v}_c)^2_\perp \right]\,.
\label{H-1PN-PM-gauge}
\end{align}
It agrees perfectly with \eqref{H-1PM-1PN-full}.

We should also check how the boost generator gets transformed.   The answer is 
\begin{align}
     \Delta \vec{G}^{(1/2)} = \{ \vec{G}^{(-1/2)}, C^{(1)} \} 
     = \frac{Gm_1m_2}{2r} \left( \zeta_1 - \zeta_2 \right) (\vec{x}_1-\vec{x}_2) \,.
\end{align}
Adding this to the $\vec{G}^{(1/2)}_\mathrm{ADM}$ in \eqref{G12-Y}, 
we obtain the boost generator in the ``PM-gauge":
\begin{align}
    \vec{G}^{(1/2)}_\mathrm{PM} = \frac{\vec{p}_1^2}{2m_1} \vec{x}_1 + \frac{\vec{p}_2^2}{2m_2} \vec{x}_2 +\left(-\frac{Gm_1m_2}{r} \right)  (\zeta_2 \vec{x}_1 + \zeta_1 \vec{x}_2 ) \,.
    \label{G12-PM-gauge}
\end{align}
The $\mathcal{O}(G)$ term agrees perfectly with \eqref{G-1PM-1PN}.

\subsection{Non-spinning 2PN}

\paragraph{2PN ADM Hamiltonian} 

The 2PN Hamiltonian in the ADM gauge is given by \cite{Damour:2000kk} 
\begin{align}
\begin{split}
    H^{(3)} &=  \frac{\vec{p}_1^6}{16 m_1^5}+ \frac{\vec{p}_2^6}{16 m_2^5} 
        \\
        &\quad + \frac{Gm_1m_2}{8r} A^{(3b)} 
        + \frac{G^2 (m_1+m_2) m_1m_2}{4r^2} A^{(3c)} - \frac{G^3m_1^2 m_2^2}{4r^3} A^{(3d)}\,.  
\end{split}
\label{H-2PN}
\end{align}
Here, the $A^{(3a),(3b),(3c)}$ are dimensionless polynomials 
of $\vec{v}_{1,2}$ and mass ratios. Explicitly, 
\begin{align}
\begin{split}
    A^{(3b)} &= 5(\vec{v}_1^4 + \vec{v}_2^4) - 11 \vec{v}_1^2 \vec{v}_2^2 -2 (\vec{v}_1\cdot\vec{v}_2)^2 -3 (\hat{n}\cdot\vec{v}_1)^2 (\hat{n}\cdot\vec{v}_2)^2 
    \\
    &\quad   +5 ( \vec{v}_1^2(\hat{n}\cdot\vec{v}_2)^2 + (\hat{n}\cdot\vec{v}_1)^2 \vec{v}_2^2)- 12 (\hat{n}\cdot\vec{v}_1)(\hat{n}\cdot\vec{v}_2)(\vec{v}_1 \cdot \vec{v}_2) \,,
    \\
    A^{(3c)} &= -27 (\vec{v}_1 \cdot\vec{v}_2) -6 (\hat{n}\cdot\vec{v}_1) (\hat{n}\cdot\vec{v}_2) 
    + 19(\zeta_1 \vec{v}_1^2 + \zeta_2 \vec{v}_2^2) + 10 (\zeta_2 \vec{v}_1^2+\zeta_1 \vec{v}_2^2)  \,,
    \\
    A^{(3d)} &= 5 + \frac{m_1}{m_2} + \frac{m_2}{m_1} \,.
\end{split}
\label{H-2PN-bcd}
\end{align}

\paragraph{2PN ADM boost} The 2PN boost, $\vec{G}^{(3/2)}$, 
first constructed in \cite{Damour:2000kk} is 
\begin{align}
\begin{split}
     \vec{G}^{(3/2)} &=  
    \sum_{a=1,2} \left( M_a \vec{x}_a + N_a \vec{v}_a \right)\,, 
    \\
    M_1 &= -\frac{1}{8} m_1 \vec{v}_1^4 + \frac{Gm_1m_2}{4r}\left[ -5\vec{v}_1^2 - \vec{v}_2^2 +7 \vec{v}_1\cdot \vec{v}_2 +(\hat{n}\cdot \vec{v}_1) (\hat{n}\cdot \vec{v}_2) \right] 
    \\
    &\hskip 2cm + \frac{G^2 m_1m_2(m_1+m_2)}{4r^2} \,,
    \\
    N_1 &= -\frac{5}{4} Gm_1m_2 (\hat{n}\cdot \vec{v}_2) \,,
\end{split}
\label{G32-ADM}
\end{align}
with $M_2$ and $N_2$ obtained by a $1\leftrightarrow 2$ relabeling
 $(\hat{n} = \hat{n}_{12} = - \hat{n}_{21})$. 
It was shown in \cite{Damour:2000kk} that \eqref{G32-ADM} 
is the unique solution to the $H$/$P$/$J$-conditions:
\begin{align}
\begin{split}
      \{ G^{(3/2)}_i , P^{(1/2)}_j \} -  \delta_{ij} H^{(2)} &= 0 \,,
    \\
\{ \vec{G}^{(3/2)} , H^{(1)} \} + 
      \{ \vec{G}^{(1/2)} , H^{(2)} \} + \{ \vec{G}^{(-1/2)} ,  H^{(3)} \} &= 0 \,, 
      \\
    \{ G^{(-1/2)}_i  , G^{(3/2)}_j \} + \{ G^{(3/2)}_i  , G^{(-1/2)}_j \} + \{ G^{(1/2)}_i  , G^{(1/2)}_j \} &= 0\,.
\end{split}
\label{G32-HPJ} 
\end{align}

\paragraph{1PM-2PN} 

To compare the PM gauge and the ADM gauge at the intersection of 1PM and 2PN, 
we should expand our 1PM Hamiltonian to the 2PN order by collecting a few terms, 
\begin{align}
\begin{split}
   H_{\mathrm{PM}}^{(3)} &= \frac{Gm_1m_2}{8r} A^{(3b)}_\mathrm{PM}
   \\
    &= H_\mathrm{b,2} +  \frac{1}{2} (\vec{v}_c)^2_\perp H_\mathrm{b,1}  + \frac{1}{8}(\vec{v}_c)^2_\perp \left[3 (\vec{v}_c)^2_\perp - 4  (\zeta _1\vec{v}_1^2 + \zeta _2\vec{v}_2^2) \right] H_\mathrm{b,0} \,,
   \\
    A^{(3b)}_\mathrm{PM} &= 5(\vec{v}_1^4 + \vec{v}_2^4) -2 \vec{v}_1^2 \vec{v}_2^2 - 16(\vec{v}_1\cdot\vec{v}_2)^2
    \\
    &\quad + (\vec{v}_c)^2_\perp \left[ -6(\vec{v}_1^2+\vec{v}_2^2) + 16 (\vec{v}_1\cdot\vec{v}_2) +4 (\zeta _1\vec{v}_1^2 + \zeta _2\vec{v}_2^2)\right] - 3(\vec{v}_c)^4_\perp \,.
\end{split}
\label{H-1PM-2PN-full}
\end{align}
Similarly, our 1PM boost generator at the 2PN order is
\begin{align}
\begin{split}
      \vec{G}_\mathrm{PM}^{(3/2)} &= \frac{Gm_1 m_2}{2r} \left[ -3 (\vec{v}_1^2+\vec{v}_2^2) + 8 \vec{v}_1 \cdot \vec{v}_2 - (\vec{v}_c)^2_\perp  \right] (\zeta_2\vec{x}_1 + \zeta_1 \vec{x}_2) 
      \\
      &\quad + \frac{Gm_1 m_2}{2r} 
      (\vec{v}_1^2 - \vec{v}_2^2) \zeta_1 \zeta_2 (\vec{x}_1 - \vec{x}_2) \,.
\end{split}
\label{G-1PM-2PN}
\end{align}

\paragraph{Canonical transformation at 2PN} 
The action of the canonical transformation at the 2PN order is given by 
\begin{align}
   \Delta H^{(3)} &= \{ H^{(1)}, C^{(2)} \} + \{ H^{(2)}, C^{(1)} \} + \frac{1}{2} \{ \{ H^{(1)}, C^{(1)} \} , C^{(1)}  \} \,,
   \label{H3-cano}
\\
\Delta \vec{G}^{(3/2)} &=  
 \{ \vec{G}^{(-1/2)} , C^{(2)} \} + 
 \{ \vec{G}^{(1/2)}_\mathrm{free} , C^{(1)} \}  
+ \frac{1}{2}  \{ \{ \vec{G}^{(-1/2)} , C^{(1)} \} , C^{(1)} \}\,.
\label{G32-cano}
\end{align}
We fixed $C^{(1)}$ earlier in \eqref{c1-ADM-PM}. 
The most general form of $C^{(2)}$ is 
\begin{align}
\begin{split}
       \frac{C^{(2)}}{Gm_1m_2} &= c_1 \vec{v}_1^2 (\hat{n}\cdot\vec{v}_1) - \bar{c}_1 \vec{v}_2^2 (\hat{n}\cdot\vec{v}_2)  
     +  c_2 \vec{v}_2^2 (\hat{n}\cdot\vec{v}_1) - \bar{c}_2 \vec{v}_1^2 (\hat{n}\cdot\vec{v}_2)  
        \\
       &\quad +  \left[ c_3  (\hat{n}\cdot\vec{v}_1) - \bar{c}_3  (\hat{n}\cdot\vec{v}_2)   \right] (\vec{v}_1\cdot\vec{v}_2) 
        \\
       &\quad +  c_4  (\hat{n}\cdot\vec{v}_1)^3 - \bar{c}_4 (\hat{n}\cdot\vec{v}_2)^3 
        +  \left[ c_5  (\hat{n}\cdot\vec{v}_1) - \bar{c}_5  (\hat{n}\cdot\vec{v}_2)   \right]  (\hat{n}\cdot\vec{v}_1) (\hat{n}\cdot\vec{v}_2)  \,.
\end{split}
\label{C2-general}
\end{align}
Again, the coefficients $c_k$, $\bar{c}_k$ are dimensionless functions of the mass ratio. The exchange symmetry requires that $c_k(m_1,m_2) = \bar{c}_k(m_2,m_1)$. 

It remains to find a set of coefficients which brings us from the ADM gauge to the PM gauge. 
In practice, it is easier to consider the shift of the boost generator \eqref{G32-cano} first.

\paragraph{Transforming the boost}

Here is a list of known facts reproduced for a reference. 
\begin{align}
\begin{split}
       \vec{G}^{(3/2)}_\mathrm{ADM} &=   
    \sum_{a=1,2} \left( M_a \vec{x}_a + N_a \vec{v}_a \right)\,, 
    \\
    &\quad M_1 = \frac{Gm_1m_2}{4r}\left[ -5\vec{v}_1^2 - \vec{v}_2^2 +7 \vec{v}_1\cdot \vec{v}_2 +(\hat{n}\cdot \vec{v}_1) (\hat{n}\cdot \vec{v}_2) \right] \,,
    \\
    &\quad N_1 = -\frac{5}{4} Gm_1m_2 (\hat{n}\cdot \vec{v}_2) \,,
    \\
    \vec{G}^{(3/2)}_\mathrm{PM} &= \frac{Gm_1 m_2}{2r} \left[ -3 (\vec{v}_1^2+\vec{v}_2^2) + 8 \vec{v}_1 \cdot \vec{v}_2 - (\vec{v}_c)^2_\perp  \right] (\zeta_2\vec{x}_1 + \zeta_1 \vec{x}_2)  
     \\
      &\quad + \frac{Gm_1 m_2}{2r} 
      (\vec{v}_1^2 - \vec{v}_2^2) \zeta_1 \zeta_2 (\vec{x}_1 - \vec{x}_2) \,,
    \\
      C^{(1)} &= \frac{1}{2} Gm_1m_2 (\zeta_1^2 \vec{v}_1 - \zeta_2^2 \vec{v}_2)  \cdot \hat{n} \,.
\end{split}
\label{G32-cano-prep}
\end{align}
We are displaying only the $\mathcal{O}(G)$ terms and suppressing all $\mathcal{O}(G^0)$ or $\mathcal{O}(G^{n\ge 2})$ terms. 
Our goal is to determine the unknown coefficients of $C^{(2)}$ as much as possible. In \eqref{G32-cano}, we first compute the known contribution, 
\begin{align}
\begin{split}
      \{ \vec{G}^{(1/2)}_\mathrm{ADM} , C^{(1)} \} &= \frac{Gm_1m_2}{2r} \left[ (\vec{v}_1 \cdot \vec{v}_2)_\perp (\zeta_2^2 \vec{x}_1 + \zeta_1^2 \vec{x}_2) - \left( \zeta_1^2 (\vec{v}_1)^2_\perp \vec{x}_1 + \zeta_2^2 (\vec{v}_2)^2_\perp \vec{x}_2 \right)  \right] 
      \\
      &\quad 
      + \frac{Gm_1m_2}{4r}  (\zeta_1^2 \vec{v}_1^2 - \zeta_2^2 \vec{v}_2^2)(\vec{x}_1 - \vec{x}_2 ) \,.
\end{split}
\label{G12-C1-cano}
\end{align}
We can do a simple consistency check. 
Schematically, $C^{(2)}$ are made of $\vec{v}^2 (\vec{v}\cdot\hat{n})$. As far as the $M_a \vec{x}_a$ terms of $\vec{G}$ are concerned, $\{ \vec{G}^{(-1/2)} , C^{(2)} \}$ can only produce terms proportional to $\hat{n} = (\vec{x}_1 -\vec{x}_2)/r$. 
It must be that 
\begin{align}
    \left[ \vec{G}^{(3/2)}_\mathrm{PM}\right]_{\vec{x}_1 = \vec{x}_2} = \left[ \vec{G}^{(3/2)}_\mathrm{ADM} 
+ \{ \vec{G}^{(1/2)}_\mathrm{free} , C^{(1)} \} \right]_{\vec{x}_1 = \vec{x}_2} \,.
\end{align}
Indeed, setting $\vec{x}_1 = \vec{x}_2 = \vec{x}$ in \eqref{G12-C1-cano}, we find 
\begin{align}
\begin{split}
  \left[ \{ \vec{G}^{(1/2)}_\mathrm{ADM} , C^{(1)} \} \right]_{\vec{x}_1 = \vec{x}_2 = \vec{x}} &= \frac{Gm_1m_2}{2M^2r} \left[ (\vec{v}_1 \cdot \vec{v}_2)_\perp (m_2^2 + m_1^2 ) - \left( (\vec{p}_1)^2_\perp  + (\vec{p}_2)^2_\perp \right)  \right] \vec{x}  
  \\
  &=\frac{Gm_1m_2}{2r} \left[ (\vec{v}_1 \cdot \vec{v}_2)_\perp -  (\vec{v}_c)^2_\perp   \right] \vec{x}  
  \\
  & = (H^{(2)}_\mathrm{PM} - H^{(2)}_\mathrm{ADM}) \vec{x} =  \left[ \vec{G}^{(3/2)}_\mathrm{PM} - \vec{G}^{(3/2)}_\mathrm{ADM} \right]_{\vec{x}_1 = \vec{x}_2 = \vec{x}} \,.
\end{split}
\end{align}
Having taken care of the overall shift of $\vec{x}_{1,2}$, 
 we may set $\vec{x}_1 = \vec{r}/2 = -\vec{x}_2$ without loss of generality. Then, the expressions in \eqref{G32-cano-prep} simplify quite a bit. 
\begin{align}
\begin{split}
       \vec{G}^{(3/2)}_\mathrm{ADM} &=-  \frac{Gm_1m_2}{2} (\vec{v}_1^2 - \vec{v}_2^2) \hat{n} - \frac{5}{4} Gm_1m_2 \left[ (\hat{n}\cdot\vec{v}_2)\vec{v}_1 - (\hat{n}\cdot\vec{v}_1)\vec{v}_2 \right]\,,
     \\
      \{ \vec{G}^{(1/2)}_\mathrm{ADM} , C^{(1)} \} &= \frac{Gm_1m_2}{4} \left[ (\vec{v}_1 \cdot \vec{v}_2)_\perp (\zeta_2^2-\zeta_1^2) + \zeta_1^2 (\hat{n}\cdot\vec{v}_1)^2 - \zeta_2 ^2 (\hat{n}\cdot\vec{v}_2)^2 \right] \hat{n} \,, 
      \\
      \vec{G}^{(3/2)}_\mathrm{PM} &= \frac{Gm_1 m_2}{4} \left[ -3 (\vec{v}_1^2+\vec{v}_2^2) + 8 \vec{v}_1 \cdot \vec{v}_2 - (\vec{v}_c)^2_\perp  \right] (\zeta_2^2-\zeta_1^2)\hat{n}    
      \\
      &\qquad + \frac{Gm_1 m_2}{4} 
      (\vec{v}_1^2 - \vec{v}_2^2) (2\zeta_1\zeta_2) \hat{n} \,.
\end{split}
\label{list-simpler}
\end{align}
In \eqref{list-simpler}, the only terms proportional to $\vec{v}_a$ reside in $\vec{G}^{(3/2)}_\mathrm{ADM}$. 
To cancel them, we need 
\begin{align}
    2c_1 + c_3 = 0 = 2\bar{c}_1 + \bar{c}_3 \,,
    \quad 
    2c_2 + c_3 = -\frac{5}{4}=  2\bar{c}_2 + \bar{c}_3\,.
\label{ck-rel-A}
\end{align}
We can use these relations to eliminate $(c_1, c_2, \bar{c}_1, \bar{c}_2)$ in favor of $(c_3, \bar{c}_3)$.
To proceed further, we read off the coefficients of monomials in 
\begin{align}
    \Delta_2 \vec{G}^{(3/2)} = \vec{G}^{(3/2)}_\mathrm{PM} - \vec{G}^{(3/2)}_\mathrm{ADM} - \{ \vec{G}^{(1/2)}_\mathrm{ADM} , C^{(1)} \} \,.
    \label{D2-G32}
\end{align}
Studying the coefficients of $\vec{v}_1^2$, $\vec{v}_2^2$ and $\vec{v}_1\cdot\vec{v}_2$, we learn that 
\begin{align}
    c_3 - \bar{c}_3 = - \frac{1}{4} (\zeta_1 -\zeta_2)(7\zeta_1^2+12\zeta_1 \zeta_2 + 7 \zeta_2^2) \,.
    \label{ck-rel-B}
\end{align}
If we had one more constraint, say for $c_3 + \bar{c}_3$, we would fix all of $(c_1,c_2,c_3,\bar{c}_1,\bar{c}_2,\bar{c}_3)$. 
But, \eqref{D2-G32} does not offer any more information. 

Studying the coefficients of $(\hat{n}\cdot\vec{v}_1)^2$, $(\hat{n}\cdot\vec{v}_2)^2$ and $(\hat{n}\cdot\vec{v}_1)(\hat{n}\cdot\vec{v}_2)$, we learn that 
\begin{align}
\begin{split}
    &3c_4 + c_5 = - \frac{1}{2} \zeta_1^3 \,,
    \quad 
    3\bar{c}_4 + \bar{c}_5 = - \frac{1}{2} \zeta_2^3 \,,
    \\
    &2(c_5 - \bar{c}_5) = - \frac{1}{4} (\zeta_1 -\zeta_2)(\zeta_1^2+4\zeta_1 \zeta_2 + \zeta_2^2) \,.
\end{split}
    \label{ck-rel-C}
\end{align}
Again, we would need one more constraint, say for $c_5 + \bar{c}_5$, to fix all of $(c_4,c_5,\bar{c}_4,\bar{c}_5)$.

\paragraph{Transforming the Hamiltonian} 

The canonical transformation \eqref{H3-cano} will shift the coefficients of $A^{(3b)}$, $A^{(3c)}$, $A^{(3d)}$ terms in \eqref{H-2PN-bcd}. For now, we are mainly interested in $A^{(3b)}$ as it is the only one relevant for the 1PM comparison.

Let us first consider the two terms involving $C^{(1)}$ in \eqref{H3-cano}. 
The last term in \eqref{H3-cano} begins at $\mathcal{O}(G^2)$, 
so it cannot affect the 1PM Hamiltonian. 
In the second term of \eqref{H3-cano}, the 1PM contribution is 
relatively simple, 
\begin{align}
\{ H^{(2)} , C^{(1)} \}_\mathrm{1PM} = \{ H^{(2a)} , C^{(1)} \} \,.
\end{align}
In terms of the $A^{(3b)}$ polynomial, the shift is 
\begin{align}
     \Delta_1 A_\mathrm{ADM} 
     =  2 \left[ \zeta_1^2 (\vec{v}_1)^2_\perp \vec{v}_1^2 + \zeta_2^2(\vec{v}_2)^2_\perp \vec{v}_2^2 - (  \vec{v}_1\cdot \vec{v}_2)_\perp (\zeta_2^2 \vec{v}_1^2 + \zeta_1^2 \vec{v}_2^2)\right] \,.
\end{align}
Recall that the PM gauge Hamiltonian is 
\begin{align}
\begin{split}
      A_\mathrm{PM}
&= 5(\vec{v}_1^4 + \vec{v}_2^4) -2 \vec{v}_1^2 \vec{v}_2^2 - 16(\vec{v}_1\cdot\vec{v}_2)^2
    \\
    &\quad + (\vec{v}_c)^2_\perp \left[ -6(\vec{v}_1^2+\vec{v}_2^2) + 16 (\vec{v}_1\cdot\vec{v}_2) +4 (\zeta _1\vec{v}_1^2 + \zeta _2\vec{v}_2^2)\right] - 3(\vec{v}_c)^4_\perp \,.
\end{split}
\label{A-PM}
\end{align}
The ADM gauge Hamiltonian is 
\begin{align}
\begin{split}
      A_\mathrm{ADM} &= 5(\vec{v}_1^4 + \vec{v}_2^4) - 11 \vec{v}_1^2 \vec{v}_2^2 -2 (\vec{v}_1\cdot\vec{v}_2)^2 -3 (\hat{n}\cdot\vec{v}_1)^2 (\hat{n}\cdot\vec{v}_2)^2 
    \\
    &\quad   +5 ( \vec{v}_1^2(\hat{n}\cdot\vec{v}_2)^2 + (\hat{n}\cdot\vec{v}_1)^2 \vec{v}_2^2)- 12 (\hat{n}\cdot\vec{v}_1)(\hat{n}\cdot\vec{v}_2)(\vec{v}_1 \cdot \vec{v}_2) \,. 
\end{split}
\label{A-ADM}
\end{align}
We wish to find a $C^{(2)}$ such that $\{ H^{(1)} , C^{(2)} \}$ gives rise to $\Delta_2 A_\mathrm{ADM}$ satisfying 
\begin{align}
  A_\mathrm{PM} =  A_\mathrm{ADM}  + \Delta_1 A_\mathrm{ADM}  + \Delta_2 A_\mathrm{ADM}  \,.
  \label{C2-final-goal}
\end{align}

In our analysis of $\Delta \vec{G}^{(3/2)}$, 
we fixed 8 out of 10 coefficients appearing in \eqref{C2-general}.  
Now, we can try to use our knowledge of $A_\mathrm{PM}$ and $A_\mathrm{ADM}$ to determine the last 2 unknowns. 
From the $\vec{v}_1^4$ coefficient, we determine $c_1$, which in turn gives 
\begin{align}
    c_3 = - \frac{1}{4} \zeta_1^2(7\zeta_1^2 + 12\zeta_1\zeta_2 + 8 \zeta_2^2) \,,
\end{align}
in accordance with \eqref{ck-rel-B}.
From the $(\hat{n}\cdot\vec{v}_1)^4$ coefficient, we determine 
$c_4$, which implies
\begin{align}
    c_5 = - \frac{1}{8} \zeta_1^3(\zeta_1+4\zeta_2) \,,
\end{align}
in agreement with the second line of \eqref{ck-rel-C}.
Using \eqref{ck-rel-A} and the first line of \eqref{ck-rel-C}, 
we fix uniquely all of the ten coefficients.

\subsection{Non-spinning 1PM}

This subsection is distinct from the previous two and the subsequent two in that 
we stay in the PM framework.
The many body 1PM ADM Hamiltonian was derived by Ledvinka, Sch\"{a}fer, and Bi\v{c}\'{a}k (LSB) in \cite{Ledvinka:2008tk}. 
When restricted to two bodies, the Hamiltonian becomes 
\begin{align}
\begin{split}
    H^{[1]}_\mathrm{LSB} = & -\frac{G E_1 E_2}{r}\left(1+\frac{\vec{p}_1^2}{E_1^2}+\frac{\vec{p}_2^2}{E_2^2} \right) +\frac{G}{2r}\left[7 (\vec{p}_1 \cdot \vec{p}_2) + (\vec{p}_1 \cdot \hat{n})(\vec{p}_2 \cdot \hat{n})\right]
    \\
    & -\frac{G}{4r}\frac{1}{E_1 E_2}\left[\frac{N_{1}}{(y_{1}+1)^2 y_{1}} + \frac{N_{2}}{(y_{2}+1)^2 y_{2}} \right] \,,
\end{split}
\label{H-LSB-original}
\end{align}
where the variables $y_a$ and $N_{ab}$ in the last line are defined by
\begin{align}
\begin{split}
y_a^2 &= 1-\frac{\vec{p}_a^2}{E_a^2} + \frac{(\vec{p}_a \cdot \hat{n})^2}{E_a^2} = \frac{m_a^2 + (\vec{p}_a\cdot\hat{n})^2}{m_a^2 + \vec{p}_a^2}\,,
\\
    N_{a} & = \frac{2}{E_b^2} \left[ 2\left(\vec{p}_1 \vec{p}_2\right)^2 \left(\vec{p}_a  \hat{n}\right)^2  -2\left(\vec{p}_1  \hat{n}\right)\left(\vec{p}_2 \hat{n}\right)\left(\vec{p}_1  \vec{p}_2\right) \vec{p}_a^2+\left(\vec{p}_{\bar{a}} \hat{n}\right)^2 \vec{p}_a^4 -\left(\vec{p}_1 \vec{p}_2\right)^2\vec{p}_a^2 \right]
    \\
    &\;\; +2\left[\left(\vec{p}_1 \hat{n}\right)^2\left(\vec{p}_2 \hat{n}\right)^2 -\vec{p}_1^2\left(\vec{p}_2 \hat{n}\right)^2
    +2\left(\vec{p}_1 \hat{n}\right)\left(\vec{p}_2  \hat{n}\right)\left(\vec{p}_1  \vec{p}_2\right)-\left(\vec{p}_1  \hat{n}\right)^2 \vec{p}_2^2 +\left(\vec{p}_1 \vec{p}_2\right)^2 \right]
    \\
    &\;\; + y_{a} \left[\left(\vec{p}_1 \hat{n}\right)^2\left(\vec{p}_2 \hat{n}\right)^2 -3 \vec{p}_1^2\left(\vec{p}_2 \hat{n}\right)^2+8\left(\vec{p}_1  \hat{n}\right)\left(\vec{p}_2 \hat{n}\right)\left(\vec{p}_1\vec{p}_2\right) -3\left(\vec{p}_1  \hat{n}\right)^2 \vec{p}_2^2 +\vec{p}_1^2 \vec{p}_2^2\right]   \,.
\end{split}
\end{align}
Here, to compress lengthy expressions, we temporarily adopt shorthand notations, 
\begin{align}
    (\vec{x}\vec{y}) = \vec{x} \cdot\vec{y} \,,
    \quad 
    \vec{p}_{\bar{1}} = \vec{p}_2 \,,
    \quad 
    \vec{p}_{\bar{2}} = \vec{p}_1 \,. 
\end{align}
The equivalence between the $H^{[1]}_\mathrm{LSB}$ and $H^{[1]}_\mathrm{PM}$ in this paper calls for a generator $C^{[1]}$,
\begin{align}
    \{H^{[0]}, C^{[1]}\} = \Delta H^{[1]} \equiv H^{[1]}_\mathrm{PM}  - H^{[1]}_\mathrm{LSB}\,.
    \label{Delta-H1}
\end{align}
Since $H^{[1]}_\mathrm{LSB}$ is known to be consistent with other PN computations up to the 4PN order \cite{Jaranowski:2012eb}, 
if we can construct $C^{[1]}$, the PN comparison in the two previous subsections will become redundant. 
Unfortunately, constructing $C^{[1]}$ in the lab frame appears to be non-trivial. 
Here, we content ourselves with an easier task: constructing $C^{[1]}$ in the COM frame.

On dimensional ground, we may $C^{[1]}$ with a dimensionless variable $c^{[1]}$, 
\begin{align}
    C^{[1]} = G E_1 E_2 \, c^{[1]} \,.
\end{align}
Since $H^{[0]}$ has no dependence on $\vec{x}_{1,2}$ and $C^{[1]}$ depends on $\vec{r}$ only through $\hat{n}$, we have 
\begin{align}
    \{H^{[0]}, C^{[1]}\} = -\frac{G E_1 E_2}{r} \left(\frac{\vec{p}_1}{E_1} - \frac{\vec{p}_2}{E_2} \right) \cdot \left(\vec{\nabla}_n c^{[1]} - \hat{n}(\hat{n} \cdot \vec{\nabla}_n c^{[1]} ) \right) \,.
    \label{cano-lab}
\end{align}
Constructing $C^{[1]}$ in the lab frame appears to be a formidable task. 
In the COM frame, where $\vec{p}=\vec{p}_1 = -\vec{p}_2$, 
the LSB Hamiltonian gets simplified drastically:
\begin{align}
\begin{split}
    H^{[1]}_\mathrm{LSB}|_\mathrm{COM} = & -\frac{G E_1 E_2}{r}\left(1+\frac{\vec{p}^2}{E_1^2}+\frac{\vec{p}^2}{E_2^2} \right) -\frac{G}{2r}\left[7 \vec{p}^2 + (\vec{p} \cdot \hat{n})^2\right]
    \\
    & -\frac{G}{4r}\frac{1}{E_1 E_2} \sum_{a=1,2} \frac{2(\vec{p}^2)^2 y_a^2 + 2(\vec{p} \cdot \hat{n})^4 +\left[(\vec{p} \cdot \hat{n})^2 +\vec{p}^2 \right]^2 y_a}{(y_{a}+1)^2 y_{a}} \,.
\end{split}
\label{H-LSB-COM}
\end{align}
Computing and simplifying $\Delta H^{[1]}$ defined in \eqref{Delta-H1}, we find 
\begin{align}
\begin{split}
    \Delta H^{[1]} =  - \frac{G}{r}  \left(\frac{\vec{p}^2-(\vec{p} \cdot \hat{n})^2}{2} \right)
    - \frac{G}{4r}\frac{1}{E_1 E_2} \sum_{a=1,2} \frac{1-y_a}{y_a} \left(2m_a^4 -E_a^4 y_a(y_a+1)\right)\,,
\end{split}
\label{Delta-H1-COM}
\end{align}
which is to be equated with \eqref{cano-lab} restricted to the COM frame, 
\begin{align}
\begin{split}
   \{H^{[0]}, C^{[1]}\}_\mathrm{COM} 
    =  -\frac{G(E_1+E_2)}{r} \left(\vec{p} \cdot \vec{\nabla}_n c^{[1]} - (\vec{p} \cdot \hat{n})(\hat{n} \cdot \vec{\nabla}_n c^{[1]} ) \right) \,.
\end{split}
\label{cano-COM}
\end{align}
In \eqref{Delta-H1-COM}, to account for the fact that $(\vec{p}\cdot\hat{n})$ and $y$ are not mutually independent, we eliminated $\vec{p}^2$ and $(\vec{p}\cdot\hat{n})^2$ in favor of $E_a^2$ and $m_a^2$. Interestingly, this rearrangement removes the cumbersome $(y+1)^2$ factors in the denominators completely. 

The first group of terms in $\Delta H^{[1]}$ stand out in that they are quadratic in $\vec{p}$; they do not mix with other terms in the PN expansion. Their contribution to $c^{[1]}$ is easy to identify:
\begin{align}
\mathcal{D} (\vec{p} \cdot \hat{n}) \equiv 
    \left(\vec{p} \cdot \vec{\nabla}_n - (\vec{p} \cdot \hat{n})(\hat{n} \cdot \vec{\nabla}_n) \right) (\vec{p} \cdot \hat{n}) 
    = \vec{p}^2-(\vec{p} \cdot \hat{n})^2 \,.
\end{align}
We may write this finding as 
\begin{align}
    c^{[1a]} = \frac{(\vec{p}\cdot \hat{n}) }{2(E_1+E_2)} 
    \quad 
    \mbox{or} 
    \quad 
    C^{[1a]} = \frac{GE_1 E_2 }{2(E_1+E_2)} (\vec{p}\cdot \hat{n}) \,.
\end{align}
To the leading PN order, $C^{[1a]}$ agrees with $C^{(1)}$ in \eqref{c1-ADM-PM} restricted to the COM frame. 

To account for the rest of $\Delta H^{[1]}$, call it $\Delta H^{[1b]}$, we begin with a simple identity, 
\begin{align}
    \mathcal{D} y_a 
    = \frac{(\vec{p} \cdot \hat{n})}{E_a^2 y_a} (\vec{p}^2-(\vec{p} \cdot \hat{n})^2) = (\vec{p} \cdot \hat{n}) \frac{1-y_a^2}{y_a} \,.
\end{align}
We use it to derive the following relation
\begin{align}
\begin{split}
    \mathcal{D} \left[ (\vec{p} \cdot \hat{n}) f(y) \right] &= \left(\vec{p}^2 - E^2(1-y^2) \right)\frac{1-y^2}{y}f'(y) + E^2(1-y^2) f(y)  
    \\
    &= \frac{1-y^2}{y}\left[ \left(E^2 y^2-m^2 \right)f'(y) + E^2y f(y) \right] \,.   
\end{split}
\end{align}
Then we can obtain $c^{[1b]}$ by solving the differential equation
\begin{align}
\begin{split}
     &\mathcal{D} \left[ (\vec{p} \cdot \hat{n}) f(y) \right] = 
    \frac{1-y}{y} \left(2m^4 -E^4 y(y+1)\right) 
    \\
    \Longrightarrow\quad 
    &(1+y) \left[ \left(E^2 y^2-m^2 \right)f'(y) + E^2y f(y) \right] = 
    2m^4 - E^4 y(y+1)   \,. 
\end{split}
\end{align}
The solution is
\begin{align}
     (\vec{p} \cdot \hat{n}) f(y) = E^2 (\vec{p} \cdot \hat{n}) -\frac{4m^4}{|\vec{p}|} \mathrm{arctanh}\left(\frac{|\vec{p}|(\vec{p} \cdot \hat{n})}{(E+m)(Ey+m)} \right) \,.
     \label{f-sol}
\end{align}
%
Combining the two contributions, we get the final form of $C^{[1]}$ in the COM frame, 
\begin{align}
\begin{split}
      C^{[1]} &= \frac{GE_1 E_2 }{2(E_1+E_2)} (\vec{p}\cdot \hat{n}) 
      \\
      &\quad +\frac{G}{4(E_1+E_2)} \sum_{a=1,2} \left[E_a^2 (\vec{p} \cdot \hat{n}) -\frac{4m_a^4}{|\vec{p}|} \mathrm{arctanh}\left(\frac{|\vec{p}|(\vec{p} \cdot \hat{n})}{(E_a+m_a)(E_a y_a+m_a)} \right) \right] \,.
\end{split}
\end{align}
%
Expanding in powers of $|\vec{p}|$,
\begin{align}
    C^{[1]} = G(\vec{p} \cdot \hat{n}) \left[\frac{m_1 m_2}{2(m_1+m_2)} + \frac{(2m_1^2+3m_1 m_2 +2m_2^2)\vec{p}^2 +m_1 m_2(\vec{p}\cdot\hat{n})^2)}{8m_1 m_2(m_1+m_2)} \right] +\mathcal{O}(p^5) \,,
\end{align}
which equals to $C^{(1)}+C^{(2)}$ from \eqref{c1-ADM-PM} and \eqref{C2-general} restricted to the COM frame.


%
%

\newpage 
\subsection{Spinning 1PN}

\paragraph{1PN with spin} 

The translation generator 
remains unchanged. 
The rotation generator $\vec{J}$ now includes spins 
as already indicated in \eqref{P-J}. It implies that 
the spin vector has zero weight in the formal PN counting. 
Nevertheless, additional powers of spin increase the PN weight of the Hamiltonian, as the spin vectors enter through 
the combination $\vec{a}_{1,2}/r$, where $\vec{a}_{1,2}$ are the spin-length vectors, $\vec{a} = \vec{S}/m$. 

The leading spin-dependent term in the Hamiltonian is the spin-orbit coupling, 
\begin{align}
    H_\mathrm{SO}^{(2)} = \frac{Gm_1m_2}{2r^2} \left[ ( 3(\vec{v}_1\times  \vec{a}_1 - \vec{v}_2\times  \vec{a}_2)  + 4(  \vec{v}_1\times  \vec{a}_2 -\vec{v}_2\times  \vec{a}_1 ) \right] \cdot \hat{n} \,. 
    \label{H2-SO-ADM}
\end{align}
Turning to the boost generators, and taking hints from \cite{Hanson:1974qy,Bel:1980ahp}, we learn that $\vec{G}^{(-1/2)}$ remains unchanged 
while $\vec{G}^{(+1/2)}$ changes slightly. 
\begin{align}
\begin{split}
      \vec{G}^{(-1/2)} &= m_1 \vec{x}_1 + m_2 \vec{x}_2  \,,
    \\
    \vec{G}^{(+1/2)} &= \frac{1}{2} \sum_{a=1,2} \left(  m_a \vec{v}_a^2 \vec{x}_a  
    + \left( -\frac{Gm_1m_2}{r}\right) \vec{x}_a 
    - \vec{S}_a \times \vec{v}_a \right)  \,. 
\end{split}
\label{G12-spin}
\end{align}
The $J$-condition holds including the spin contribution, 
\begin{align}
    \{ G^{(-1/2)}_i , G^{(1/2)}_j \} +  \{ G^{(1/2)}_i , G^{(-1/2)}_j \} = - \epsilon_{ijk} J_k \,.
\end{align}
The spin contributions to the $P$-condition all cancel out: 
\begin{align}
    \{ \vec{G}^{(1/2)}_\mathrm{spin} , H^{(1)} \} + \{ \vec{G}^{(-1/2)} ,  H^{(2)}_\mathrm{SO}  \} = 0 \,. 
    \label{G12-P-again}
\end{align}
So, no further change is needed for $\vec{G}^{(1/2)}$.

\paragraph{Canonical transformation at 1PN with spin} 
At the leading order, the possibilities for canonical transformation allowed by the PN power-counting and $\Delta H^{(2)}_\mathrm{SO} \propto r^{-2}$  are
\begin{align}
    C^{(1)}_\mathrm{SO} = \frac{Gm_1 m_2}{r}\left[b_1 \vec{a}_1 - \bar{b}_1 \vec{a}_2 \right] \cdot \hat{n} + M \left(\vec{v}_1 \times \vec{v}_2 \right) \cdot (b_2 \vec{a}_1 - \bar{b}_2 \vec{a}_2) \,,
    \label{C1-SO} 
\end{align}
where ($b_k$, $\bar{b}_k$) are dimensionless functions of the mass ratio. 
The resulting shift of the Hamiltonian is 
\begin{align}
\begin{split}
    \Delta H^{(2)}_\mathrm{SO} & = \frac{Gm_1m_2}{r^2} \left[ ( b_1 \vec{a}_1  -  \bar{b}_1 \vec{a}_2 )\cdot \hat{n}\right]  
    \left[ ( \vec{v}_1  - \vec{v}_2 )\cdot \hat{n} \right]
    +  \frac{GM^2 }{r^2} ( b_2 \vec{a}_1 - \bar{b}_2  \vec{a}_2 ) \cdot (\hat{n} \times \vec{v}_c) \,.
\end{split}
\end{align}

\paragraph{PN expansion of 1PM-SO } 

When we expand the spin-linear Hamiltonian \eqref{H1-SO-all} in the PN way, only $H^{[1a]}$ and $H^{[1c]}$ contribute at the 1PN order. The result is  
\begin{align}
      H^{(2)}_\mathrm{SO-PM} &= \frac{G m_1 m_2}{r^2}\left[2(\vec{v}_1  - \vec{v}_2 )\times (\vec{a}_1 +\vec{a}_2) - \frac{1}{2} (\vec{v}_1 \times \vec{a}_1) + \frac{1}{2}(\vec{v}_2 \times \vec{a}_2) 
  \right] \cdot \hat{n}   \,.
  \label{H2-SO-PM}
\end{align}
It agrees perfectly with \eqref{H2-SO-ADM}. There is no need for a canonical transformation. 
If $C^{(1)}_\mathrm{SO}$ were non-vanishing, 
it would induce the shift $\Delta \vec{G}^{(1/2)} = \{ \vec{G}^{(-1/2)} , C^{(1)}_\mathrm{SO} \}$, 
which would contradict the \emph{free} nature of $\vec{G}^{(1/2)}$.
%

\subsection{Spinning 2PN} 

In our PN counting, NLO SO amounts to 2PN (weight 3). 
In what follows, we will use NLO SO and 2PN SO interchangeably. 

\paragraph{NLO SO ADM} 

The NLO SO Hamiltonian in the ADM gauge is given by \cite{Damour:2007nc}.
\begin{align}
\begin{split}
      H^{(3)}_\mathrm{SO} &=  \frac{Gm_1m_2}{8r^2}
      \left[ (\vec {a}_{1} \times \hat{n}) \cdot \vec{v}_1 \left[-5\vec{v}_1^2 + 6\vec{v}_2^2 -6 \vec{v}_1 \cdot \vec{v}_2   - 12 (\vec{v}_2)^2_\parallel -6  (\vec{v}_1 \cdot \vec{v}_2)_\parallel \right]\right. 
      \\
&\hskip 3cm
+ 8 (\vec{a}_1 \times \hat{n}) \cdot \vec{v}_2 \left[ \vec{v}_1 \cdot \vec{v}_2+3 (\vec{v}_1 \cdot \vec{v}_2)_\parallel \right] 
\\
&\hskip 3cm 
 +2(\vec{v}_1 \times \vec{v}_2)\cdot \vec{a}_1 \left[3 (\hat{n} \cdot \vec{v}_1) -8 (\hat{n} \cdot \vec{v}_2) \right] \Big]
 \\
 &\qquad 
 +\frac{G^2m_1m_2}{2r^3} (\vec{a}_1 \times \hat{n}) \cdot\left[3 \vec{v}_2\left(4 m_1 +5 m_2\right)-\vec{v}_1\left(11 m_1 +10 m_2\right)\right]
  \\
 &\quad + (1 \longleftrightarrow 2) \,.
\end{split}
\end{align}
The boost generator is given by 
\begin{align}
\begin{split}
    \vec{G}^{(3/2)}_\mathrm{SO} &=  \frac{1}{8} \vec{v}_1^2(\vec{S}_1 \times \vec{v}_1)
+\frac{G m_1 m_2}{2r^2} (\vec{a}_1\times \hat{n})\cdot (3\vec{v}_1-4 \vec{v}_2)   \vec{x}_1
 \\
&\quad +\frac{G m_1m_2}{4r}\left[ \vec{a}_1 \times (5  \vec{v}_1-6 \vec{v}_2 ) -2(\hat{n} \cdot  \vec{v}_2) (\hat{n} \times \vec{a}_1)
- (\vec{a}_1\times \hat{n})\cdot (\vec{v}_1 -4 \vec{v}_2) \hat{n}\right] 
\\
&\quad + (1 \longleftrightarrow 2)\,.
\end{split}
\label{G32-SO-full}
\end{align}
It is straightforward (but lengthy) to verify the SO parts of the $H$/$P$/$J$-conditions: 
\begin{align}
\begin{split}
      \{ G^{(3/2)}_i , P^{(1/2)}_j \} -  \delta_{ij} H^{(2)} &= 0 \,,
    \\
\{ \vec{G}^{(3/2)} , H^{(1)} \} + 
      \{ \vec{G}^{(1/2)} , H^{(2)} \} + \{ \vec{G}^{(-1/2)} ,  H^{(3)} \} &= 0 \,, 
      \\
    \{ G^{(-1/2)}_i  , G^{(3/2)}_j \} + \{ G^{(3/2)}_i  , G^{(-1/2)}_j \} + \{ G^{(1/2)}_i  , G^{(1/2)}_j \} &= 0\,.
\end{split}
\label{G32-HPJ-SO} 
\end{align}

\paragraph{Transforming the boost, part 1} 

The spin-orbit contribution to the boost generator in the PM gauge is, 
in terms of $H^{(2)}_\mathrm{SO-PM}$ in \eqref{H2-SO-PM}, 
\begin{align}
    \vec{G}^{(3/2)}_\mathrm{SO-PM} = H^{(2)}_\mathrm{SO-PM} \left( \zeta_2 \vec{x}_1 + \zeta_1 \vec{x}_2 \right) - \frac{Gm_1m_2}{2r} \left[ \zeta_2(\vec{a}_1 \times \vec{v}_1) + \zeta_1 (\vec{a}_2 \times \vec{v}_2)\right] \,, 
    \label{G32-SO-PM}
\end{align}
Taking the difference between the 1PM part of \eqref{G32-SO-full} 
and \eqref{G32-SO-PM}, we find 
\begin{align}
\begin{split}
        \Delta \vec{G}^{(3/2)} &= \vec{G}^{(3/2)}_\mathrm{PM} - \vec{G}^{(3/2)}_\mathrm{ADM}
        \\
        &= - \frac{Gm_1m_2}{2r} \hat{n}\cdot \left[ \zeta_1 (3\vec{v}_1-4\vec{v}_2) \times \vec{a}_1 - \zeta_2 (3\vec{v}_2-4\vec{v}_1) \times \vec{a}_2 \right] \hat{n}
        \\
        &\quad + \frac{G m_1m_2}{4r} \hat{n}\cdot \left[ (\vec{v}_1 -4\vec{v}_2)\times \vec{a}_1 - (\vec{v}_2 -4\vec{v}_1)\times \vec{a}_2 \right] \hat{n}
        \\
        &\quad - \frac{Gm_1m_2}{2r} \left[ \zeta_2(\vec{a}_1 \times \vec{v}_1) + \zeta_1 (\vec{a}_2 \times \vec{v}_2)\right]
        \\
        &\quad -\frac{G m_1m_2}{4r}\left[ \vec{a}_1 \times (5  \vec{v}_1-6 \vec{v}_2 ) -2(\hat{n} \cdot  \vec{v}_2) (\hat{n} \times \vec{a}_1) \right] 
        \\
         &\quad -\frac{G m_1m_2}{4r}\left[ \vec{a}_2 \times (5  \vec{v}_2-6 \vec{v}_1 ) -2(\hat{n} \cdot  \vec{v}_1) (\hat{n} \times \vec{a}_2) \right] \,.
\end{split}
\end{align}
Upon a canonical transformation, the shift of $\vec{G}^{(3/2)}_\mathrm{SO}$ is given by 
\begin{align}
    \Delta \vec{G}^{(3/2)}_\mathrm{SO} = \{ \vec{G}^{(-1/2)} , C^{(2)}_\mathrm{SO} \} + \{ \vec{G}^{(1/2)} , C^{(1)}_\mathrm{SO} \} + \{ \vec{G}^{(1/2)}_\mathrm{spin} , C^{(1)}_\mathrm{no-spin} \} \,.
    \label{G32-SO-shift}
\end{align}
We learned in \eqref{H2-SO-PM} that $C^{(1)}_\mathrm{SO} = 0$, 
so the second term in \eqref{G32-SO-shift} vanishes. 
As for the third term in \eqref{G32-SO-shift}, 
using \eqref{G12-spin} and \eqref{c1-ADM-PM}, we find 
\begin{align}
\begin{split}
     \{ \vec{G}^{(1/2)}_\mathrm{spin} , C^{(1)}_\mathrm{no-spin} \} &= \frac{Gm_1m_2}{4r} \left[ (\vec{a}_1-\vec{a}_2) \times (\zeta_1^2 \vec{v}_1 - \zeta_2^2 \vec{v}_2) 
     \right.
     \\
     &\hskip 3cm \left. +[\hat{n} \times (\vec{a}_1-\vec{a}_2) ] \hat{n}\cdot(\zeta_1^2 \vec{v}_1 - \zeta_2^2 \vec{v}_2) \right]\,.
\end{split}
\label{G12s-C1ns}
\end{align}

\paragraph{Transforming the boost, part 2} 

The general form of $C^{(2)}_\mathrm{SO}$ is 
\begin{align}
\begin{split}
    C^{(2)}_\mathrm{SO} &= \frac{Gm_1m_2}{4r} P(a,v,n) \,,
    \\
    P&= P_1 + P_2 + P_3 + P_4 + P_5 
    \\
    &= \vec{a}_1\cdot \left[  (\vec{v}_1\times \hat{n}) (c_1 \vec{v}_1 + c_2 \vec{v}_2)\cdot \hat{n} + (\vec{v}_2\times \hat{n}) (c_3 \vec{v}_1 + c_4 \vec{v}_2)\cdot \hat{n} + c_5 (\vec{v}_1\times \vec{v}_2) \right] 
    \\
    &\quad +\vec{a}_2\cdot \left[  (\vec{v}_2\times \hat{n}) (\bar{c}_1 \vec{v}_2 + \bar{c}_2 \vec{v}_1)\cdot \hat{n} + (\vec{v}_1\times \hat{n}) (\bar{c}_3 \vec{v}_2 + \bar{c}_4 \vec{v}_1)\cdot \hat{n} +\bar{c}_5 (\vec{v}_2\times \vec{v}_1) \right] \,.
\end{split}
\end{align}
The shift of $\vec{G}^{(3/2)}_\mathrm{SO}$ generated by $C^{(2)}_\mathrm{SO}$ is then
\begin{align}
\begin{split}
      \{ \vec{G}^{(-1/2)} , C^{(2)}_\mathrm{SO} \} &= 
      \frac{Gm_1m_2}{4r} (\vec{Q}_1 +\vec{Q}_2 +\vec{Q}_3 + \vec{Q}_4 +\vec{Q}_5) \,,
   \\
    \vec{Q}_1 &= c_1 \left[ \vec{a}_1 \cdot (\vec{v}_1\times \hat{n})\hat{n} + (\hat{n}\times \vec{a}_1) (\hat{n}\cdot\vec{v}_1) \right] 
    \\
    &\qquad + \bar{c}_1 \left[ \vec{a}_2 \cdot (\vec{v}_2\times \hat{n})\hat{n} + (\hat{n}\times \vec{a}_2) (\hat{n}\cdot\vec{v}_2) \right] \,,
    \\
    \vec{Q}_2 &= c_2 \left[ \vec{a}_1 \cdot (\vec{v}_1\times \hat{n})\hat{n} + (\hat{n}\times \vec{a}_1) (\hat{n}\cdot\vec{v}_2) \right] 
    \\
    &\qquad + \bar{c}_2 \left[ \vec{a}_2 \cdot (\vec{v}_2\times \hat{n})\hat{n} + (\hat{n}\times \vec{a}_2) (\hat{n}\cdot\vec{v}_1) \right] \,,
    \\
    \vec{Q}_3 &= c_3 \left[ \vec{a}_1 \cdot (\vec{v}_2\times \hat{n})\hat{n} + (\hat{n}\times \vec{a}_1) (\hat{n}\cdot\vec{v}_1) \right] 
    \\
    &\qquad + \bar{c}_3 \left[ \vec{a}_2 \cdot (\vec{v}_1\times \hat{n})\hat{n} + (\hat{n}\times \vec{a}_2) (\hat{n}\cdot\vec{v}_2) \right] \,,
    \\
    \vec{Q}_4 &= c_4 \left[ \vec{a}_1 \cdot (\vec{v}_2\times \hat{n})\hat{n} + (\hat{n}\times \vec{a}_1) (\hat{n}\cdot\vec{v}_2) \right] 
    \\
    &\qquad + \bar{c}_4 \left[ \vec{a}_2 \cdot (\vec{v}_1\times \hat{n})\hat{n} + (\hat{n}\times \vec{a}_2) (\hat{n}\cdot\vec{v}_1) \right] \,,
    \\
    \vec{Q}_5 &=  (c_5\vec{a}_1 -\bar{c}_5 \vec{a}_2) \times (\vec{v}_1- \vec{v}_2) \,.
\end{split}
\end{align}
Clearly, $\vec{Q}_5$ is quite distinct from the other four. 
We find it convenient to reorganize the similar four as 
\begin{align}
\begin{split}
        \vec{Q}_1 + \vec{Q}_2 +\vec{Q}_3 +\vec{Q}_4 &= 
    \hat{n} \left[ (c_1 +c_2) (a_1,v_1,n) + (c_3+c_4) (a_1,v_2,n) \right] 
     \\
    &\quad 
    +\hat{n} \left[ (\bar{c}_1 +\bar{c}_2) (a_2,v_2,n) + (\bar{c}_3+\bar{c}_4) (a_2,v_1,n) \right] 
    \\
    &\quad + (\hat{n}\times \vec{a}_1) \left[ (c_1 +c_3) (\hat{n}\cdot\vec{v}_1)  + (c_2 +c_4) (\hat{n}\cdot\vec{v}_2) \right]
    \\
    &\quad + (\hat{n}\times \vec{a}_2) \left[ (\bar{c}_1 +\bar{c}_3) (\hat{n}\cdot\vec{v}_2)  + (\bar{c}_2 +\bar{c}_4) (\hat{n}\cdot\vec{v}_1) \right] \,.
\end{split}
\end{align}
As before, we expect to be able to fix most (but not all) coefficients by matching $\vec{G}^{(3/2)}$ between ADM and PM. 
The remaining ones can be fixed by matching $H^{(3)}$.

From the $(\vec{a}\times \vec{v})$ terms, 
we find 
\begin{align}
    c_5 + \zeta_1^2 = -5 - 2\zeta_2 \,,
    \quad 
    - c_5 - \zeta_2^2 = 6 \,.
\end{align}
The two equations are consistent with $\zeta_1 + \zeta_2 = 1$. 
The solution is 
\begin{align}
    c_5 = - (6\zeta_1^2 + 12\zeta_1 \zeta_2 + 7\zeta_2^2) \,,
    \quad 
    \bar{c}_5 = - (7\zeta_1^2 + 12\zeta_1 \zeta_2 + 6\zeta_2^2) \,.
\end{align}
From the $(\hat{n}\times \vec{a})(\hat{n}\cdot\vec{v})$ terms, we find 
\begin{align}
    c_1 + c_3 + \zeta_1^2 = 0\,,
    \quad 
    c_2 + c_4 - \zeta_2^2 = 2 \,.
    \label{c1324}
\end{align}
From the $[\hat{n}\cdot(\vec{v}\times \vec{a})]\hat{n}$ terms, 
we find 
\begin{align}
    c_1 + c_2 = 6\zeta_1 - 1 \,,
    \quad 
    c_3 + c_4 = -8\zeta_1 + 4 \,.
    \label{c1234}
\end{align}
The two sets of equations, \eqref{c1324} and \eqref{c1234}, 
agree on the sum: $c_1 + c_2 + c_3 + c_4 = \zeta_1 + 3 \zeta_2$. 

\paragraph{Transforming the Hamiltonian}

The NLO SO Hamiltonian  is given by 
\begin{align}
\begin{split}
    H^{[1](3)}_\mathrm{SO} &= \frac{Gm_1m_2}{8r^2} \sum_{a=1,2} A^{(3)}_a  \,,
    \\
    A^{(3)}_1 &= (a_1,v_1,n) \left[5\vec{v}_1^2 +8\vec{v}_1\cdot\vec{v}_2 -2\vec{v}_2^2 -6\vec{v}_c^2 +18(\hat{n}\cdot\vec{v}_c)^2 \right] 
    \\
    &\quad +(a_1,v_2 ,n) \left[ -16\vec{v}_1 \cdot \vec{v}_2 +8 \vec{v}_c^2 -24(\hat{n}\cdot \vec{v}_c)^2 \right]  
    \\
    &\qquad + (a_1,v_1,v_2) \left[ (4\zeta_1\zeta_2 - 8) (\hat{n}\cdot\vec{v}_1) + 4 \zeta_2^2(\hat{n}\cdot\vec{v}_2) \right] \,.
\end{split}
\end{align}
See appendix~\ref{sec:more-PN} for the derivation of this expression. 

Recall that the shift of $H^{(3)}$ is 
\begin{align}
    \Delta H^{(3)} = \{ H^{(1)} , C^{(2)} \} + \{ H^{(2)} , C^{(1)} \} + \frac{1}{2} \{ \{H^{(1)} , C^{(1)} \} , C^{(1)} \}  \,.
    \nonumber 
\end{align}
Ignoring all $\mathcal{O}(G^2)$ contributions, we need only to compute the first term:
\begin{align}
\begin{split}
      \{ H^{(1)} , C^{(2)}_\mathrm{SO} \} &= 
    \frac{Gm_1m_2}{4r^2} \left( R_1 +R_2 +R_3 +R_4 +R_5\right)\,,
    \\
    R_1 &= (\vec{v}_1 -\vec{v}_2)\cdot \hat{n} P_1 - c_1 \left[ (a_1,v_1,w_{12}) (\hat{n}\cdot \vec{v}_1) + (a_1,v_1,n) (\vec{w}_{12}\cdot\vec{v}_1) \right]
    \\
    &\hskip 3cm - \bar{c}_1 \left[ (a_2,v_2,w_{12}) (\hat{n}\cdot \vec{v}_2) + (a_2,v_2,n) (\vec{w}_{12}\cdot\vec{v}_2) 
 \right] \,, 
    \\
    R_2 &= (\vec{v}_1 -\vec{v}_2)\cdot \hat{n} P_2 - c_2 \left[ (a_1,v_1,w_{12}) (\hat{n}\cdot \vec{v}_2) + (a_1,v_1,n) (\vec{w}_{12}\cdot\vec{v}_2) \right]
    \\
    &\hskip 3cm - \bar{c}_2 \left[ (a_2,v_2,w_{12}) (\hat{n}\cdot \vec{v}_1) + (a_2,v_2,n) (\vec{w}_{12}\cdot\vec{v}_1) 
 \right] \,, 
 \\
    R_3 &= (\vec{v}_1 -\vec{v}_2)\cdot \hat{n} P_3 - c_3 \left[ (a_1,v_2,w_{12}) (\hat{n}\cdot \vec{v}_1) + (a_1,v_2,n) (\vec{w}_{12}\cdot\vec{v}_1) \right]
    \\
    &\hskip 3cm - \bar{c}_3 \left[ (a_2,v_1,w_{12}) (\hat{n}\cdot \vec{v}_2) + (a_2,v_1,n) (\vec{w}_{12}\cdot\vec{v}_2) 
 \right] \,, 
 \\
    R_4 &= (\vec{v}_1 -\vec{v}_2)\cdot \hat{n} P_4 - c_4 \left[ (a_1,v_2,w_{12}) (\hat{n}\cdot \vec{v}_2) + (a_1,v_2,n) (\vec{w}_{12}\cdot\vec{v}_2) \right]
    \\
    &\hskip 3cm - \bar{c}_4 \left[ (a_2,v_1,w_{12}) (\hat{n}\cdot \vec{v}_1) + (a_2,v_1,n) (\vec{w}_{12}\cdot\vec{v}_1) 
 \right] \,, 
 \\
    R_5 &= (\vec{v}_1 -\vec{v}_2)\cdot \hat{n} P_5 = (\vec{v}_1 -\vec{v}_2)\cdot \hat{n} \left( c_5 \vec{a}_1 - \bar{c}_5 \vec{a}_2 \right)\cdot (\vec{v}_1 \times \vec{v}_2) \,,
    \\
    &\quad \vec{w}_{12} = (\vec{v_1} - \vec{v}_2)_\perp = (\vec{v_1} - \vec{v}_2) - [(\vec{v_1} - \vec{v}_2)\cdot\hat{n}] \hat{n} \,.
\end{split}
\end{align}
By matching (a subset of) the coefficients of the $\Delta H^{(3)}$, we obtain our final answer for the canonical transformation:
\begin{align}
    c_1 = 3\zeta_1^2 \,,
    \quad
    c_2 = 2- 3\zeta_2^2 \,,
    \quad 
    c_3 = -4\zeta_1^2 \,,
    \quad 
    c_4 = 4\zeta_2^2 \,.
    \label{C2-SO-final}
\end{align}

\newpage 
\bibliographystyle{JHEP}
\bibliography{reference}

@article{Bjerrum-Bohr:2022blt,
    author = "Bjerrum-Bohr, N. E. J. and Damgaard, P. H. and Plante, L. and Vanhove, P.",
    title = "{The SAGEX review on scattering amplitudes Chapter 13: Post-Minkowskian expansion from scattering amplitudes}",
    eprint = "2203.13024",
    archivePrefix = "arXiv",
    primaryClass = "hep-th",
    reportNumber = "IPhT-t22/03, SAGEX-22-14",
    doi = "10.1088/1751-8121/ac7a78",
    journal = "J. Phys. A",
    volume = "55",
    number = "44",
    pages = "443014",
    year = "2022"
}

@article{Kosower:2022yvp,
    author = "Kosower, David A. and Monteiro, Ricardo and O'Connell, Donal",
    title = "{The SAGEX review on scattering amplitudes Chapter 14: Classical gravity from scattering amplitudes}",
    eprint = "2203.13025",
    archivePrefix = "arXiv",
    primaryClass = "hep-th",
    reportNumber = "SAGEX-22-15, QMUL-PH-22-08",
    doi = "10.1088/1751-8121/ac8846",
    journal = "J. Phys. A",
    volume = "55",
    number = "44",
    pages = "443015",
    year = "2022"
}

@article{Kalin:2020mvi,
    author = {K\"alin, Gregor and Porto, Rafael A.},
    title = "{Post-Minkowskian Effective Field Theory for Conservative Binary Dynamics}",
    eprint = "2006.01184",
    archivePrefix = "arXiv",
    primaryClass = "hep-th",
    reportNumber = "DESY20-077, SLAC-PUB-17529",
    doi = "10.1007/JHEP11(2020)106",
    journal = "JHEP",
    volume = "11",
    pages = "106",
    year = "2020"
}

@article{Arkani-Hamed:2017jhn,
    author = "Arkani-Hamed, Nima and Huang, Tzu-Chen and Huang, Yu-tin",
    title = "{Scattering amplitudes for all masses and spins}",
    eprint = "1709.04891",
    archivePrefix = "arXiv",
    primaryClass = "hep-th",
    reportNumber = "NCTS-TH/1714, NCTS-TH-1714",
    doi = "10.1007/JHEP11(2021)070",
    journal = "JHEP",
    volume = "11",
    pages = "070",
    year = "2021"
}

@article{Chung:2020rrz,
    author = "Chung, Ming-Zhi and Huang, Yu-tin and Kim, Jung-Wook and Lee, Sangmin",
    title = "{Complete Hamiltonian for spinning binary systems at first post-Minkowskian order}",
    eprint = "2003.06600",
    archivePrefix = "arXiv",
    primaryClass = "hep-th",
    reportNumber = "NCTS-TH/2004",
    doi = "10.1007/JHEP05(2020)105",
    journal = "JHEP",
    volume = "05",
    pages = "105",
    year = "2020"
}

@article{Bern:2021dqo,
    author = "Bern, Zvi and Parra-Martinez, Julio and Roiban, Radu and Ruf, Michael S. and Shen, Chia-Hsien and Solon, Mikhail P. and Zeng, Mao",
    title = "{Scattering Amplitudes and Conservative Binary Dynamics at ${\cal O}(G^4)$}",
    eprint = "2101.07254",
    archivePrefix = "arXiv",
    primaryClass = "hep-th",
    reportNumber = "CALT-TH-2021-004, FR-PHENO-2021-03, OUTP-21-03P",
    doi = "10.1103/PhysRevLett.126.171601",
    journal = "Phys. Rev. Lett.",
    volume = "126",
    number = "17",
    pages = "171601",
    year = "2021"
}

@article{Bern:2021yeh,
    author = "Bern, Zvi and Parra-Martinez, Julio and Roiban, Radu and Ruf, Michael S. and Shen, Chia-Hsien and Solon, Mikhail P. and Zeng, Mao",
    title = "{Scattering Amplitudes, the Tail Effect, and Conservative Binary Dynamics at $\mathcal{O}(G^4)$}",
    eprint = "2112.10750",
    archivePrefix = "arXiv",
    primaryClass = "hep-th",
    doi = "10.1103/PhysRevLett.128.161103",
    journal = "Phys. Rev. Lett.",
    volume = "128",
    number = "16",
    pages = "161103",
    year = "2022"
}

@article{Damour:2020tta,
    author = "Damour, Thibault",
    title = "{Radiative contribution to classical gravitational scattering at the third order in $G$}",
    eprint = "2010.01641",
    archivePrefix = "arXiv",
    primaryClass = "gr-qc",
    doi = "10.1103/PhysRevD.102.124008",
    journal = "Phys. Rev. D",
    volume = "102",
    number = "12",
    pages = "124008",
    year = "2020"
}

@article{Dlapa:2022lmu,
    author = {Dlapa, Christoph and K\"alin, Gregor and Liu, Zhengwen and Neef, Jakob and Porto, Rafael A.},
    title = "{Radiation Reaction and Gravitational Waves at Fourth Post-Minkowskian Order}",
    eprint = "2210.05541",
    archivePrefix = "arXiv",
    primaryClass = "hep-th",
    doi = "10.1103/PhysRevLett.130.101401",
    journal = "Phys. Rev. Lett.",
    volume = "130",
    number = "10",
    pages = "101401",
    year = "2023"
}

@inproceedings{Buonanno:2022pgc,
    author = "Buonanno, Alessandra and Khalil, Mohammed and O'Connell, Donal and Roiban, Radu and Solon, Mikhail P. and Zeng, Mao",
    title = "{Snowmass White Paper: Gravitational Waves and Scattering Amplitudes}",
    booktitle = "{2022 Snowmass Summer Study}",
    eprint = "2204.05194",
    archivePrefix = "arXiv",
    primaryClass = "hep-th",
    month = "4",
    year = "2022"
}

@article{Levi:2022dqm,
    author = "Levi, Mich\`ele and Morales, Roger and Yin, Zhewei",
    title = "{From the EFT of Spinning Gravitating Objects to Poincar\'e and Gauge Invariance}",
    eprint = "2210.17538",
    archivePrefix = "arXiv",
    primaryClass = "hep-th",
    reportNumber = "UUITP-48/22",
    month = "10",
    year = "2022"
}

@article{Levi:2022rrq,
    author = "Levi, Mich\`ele and Yin, Zhewei",
    title = "{Completing the Fifth PN Precision Frontier via the EFT of Spinning Gravitating Objects}",
    eprint = "2211.14018",
    archivePrefix = "arXiv",
    primaryClass = "hep-th",
    reportNumber = "UUITP-54/22",
    month = "11",
    year = "2022"
}

@article{Damour:2000kk,
    author = "Damour, Thibault and Jaranowski, Piotr and Schaefer, Gerhard",
    title = "{Poincare invariance in the ADM Hamiltonian approach to the general relativistic two-body problem}",
    eprint = "gr-qc/0003051",
    archivePrefix = "arXiv",
    doi = "10.1103/PhysRevD.62.021501",
    journal = "Phys. Rev. D",
    volume = "62",
    pages = "021501",
    year = "2000",
    note = "[Erratum: Phys.Rev.D 63, 029903 (2001)]"
}

@article{Damour:2007nc,
    author = "Damour, Thibault and Jaranowski, Piotr and Schaefer, Gerhard",
    title = "{Hamiltonian of two spinning compact bodies with next-to-leading order gravitational spin-orbit coupling}",
    eprint = "0711.1048",
    archivePrefix = "arXiv",
    primaryClass = "gr-qc",
    doi = "10.1103/PhysRevD.77.064032",
    journal = "Phys. Rev. D",
    volume = "77",
    pages = "064032",
    year = "2008"
}

@article{Steinhoff:2008zr,
    author = "Steinhoff, Jan and Schaefer, Gerhard and Hergt, Steven",
    title = "{ADM canonical formalism for gravitating spinning objects}",
    eprint = "0805.3136",
    archivePrefix = "arXiv",
    primaryClass = "gr-qc",
    doi = "10.1103/PhysRevD.77.104018",
    journal = "Phys. Rev. D",
    volume = "77",
    pages = "104018",
    year = "2008"
}

@article{Hergt:2008jn,
    author = "Hergt, Steven and Schaefer, Gerhard",
    title = "{Higher-order-in-spin interaction Hamiltonians for binary black holes from Poincare invariance}",
    eprint = "0809.2208",
    archivePrefix = "arXiv",
    primaryClass = "gr-qc",
    doi = "10.1103/PhysRevD.78.124004",
    journal = "Phys. Rev. D",
    volume = "78",
    pages = "124004",
    year = "2008"
}

@article{Hartung:2013dza,
    author = "Hartung, Johannes and Steinhoff, Jan and Schafer, Gerhard",
    title = "{Next-to-next-to-leading order post-Newtonian linear-in-spin binary Hamiltonians}",
    eprint = "1302.6723",
    archivePrefix = "arXiv",
    primaryClass = "gr-qc",
    doi = "10.1002/andp.201200271",
    journal = "Annalen Phys.",
    volume = "525",
    pages = "359--394",
    year = "2013"
}

@article{Levi:2016ofk,
    author = "Levi, Michele and Steinhoff, Jan",
    title = "{Complete conservative dynamics for inspiralling compact binaries with spins at the fourth post-Newtonian order}",
    eprint = "1607.04252",
    archivePrefix = "arXiv",
    primaryClass = "gr-qc",
    doi = "10.1088/1475-7516/2021/09/029",
    journal = "JCAP",
    volume = "09",
    pages = "029",
    year = "2021"
}

@article{Hanson:1974qy,
    author = "Hanson, Andrew J. and Regge, T.",
    title = "{The Relativistic Spherical Top}",
    reportNumber = "Print-74-0833 (IAS,PRINCETON)",
    doi = "10.1016/0003-4916(74)90046-3",
    journal = "Annals Phys.",
    volume = "87",
    pages = "498",
    year = "1974"
}

@article{Bel:1980ahp,
     author = {Bel, L. and Martin, J.},
     title = {Predictive relativistic mechanics of systems of {N} particles with spin},
     journal = {Annales de l'institut Henri Poincar\'e. Section A, Physique Th\'eorique},
     pages = {409--442},
     publisher = {Gauthier-Villars},
     volume = {33},
     number = {4},
     year = {1980},
     mrnumber = {605200},
     language = {en},
     url = {http://www.numdam.org/item/AIHPA_1980__33_4_409_0/}
}

@article{Schafer:2018kuf,
    author = {Sch\"afer, Gerhard and Jaranowski, Piotr},
    title = "{Hamiltonian formulation of general relativity and post-Newtonian dynamics of compact binaries}",
    eprint = "1805.07240",
    archivePrefix = "arXiv",
    primaryClass = "gr-qc",
    doi = "10.1007/s41114-018-0016-5",
    journal = "Living Rev. Rel.",
    volume = "21",
    number = "1",
    pages = "7",
    year = "2018"
}

@article{Aoude:2023vdk,
    author = "Aoude, Rafael and Haddad, Kays and Helset, Andreas",
    title = "{Classical gravitational scattering at $\mathcal{O}(G^{2} S_{1}^{\infty} S_{2}^{\infty})$}",
    eprint = "2304.13740",
    archivePrefix = "arXiv",
    primaryClass = "hep-th",
    reportNumber = "CP3-23-19, UUITP-10/23, CALT-TH-2023-010",
    month = "4",
    year = "2023"
}

@article{Kim:2021rda,
    author = "Kim, Joon-Hwi and Kim, Jung-Wook and Lee, Sangmin",
    title = "{The relativistic spherical top as a massive twistor}",
    eprint = "2102.07063",
    archivePrefix = "arXiv",
    primaryClass = "hep-th",
    reportNumber = "QMUL-PH-21-08",
    doi = "10.1088/1751-8121/ac11be",
    journal = "J. Phys. A",
    volume = "54",
    number = "33",
    pages = "335203",
    year = "2021"
}

@article{Steinhoff:2015ksa,
    author = "Steinhoff, Jan",
    title = "{Spin gauge symmetry in the action principle for classical relativistic particles}",
    eprint = "1501.04951",
    archivePrefix = "arXiv",
    primaryClass = "gr-qc",
    month = "1",
    year = "2015"
}

@article{Georg:2015afa,
    author = {Georg, Ira and Sch\"afer, Gerhard},
    title = "{Canonical center and relative coordinates for compact binary systems through second post-Newtonian order}",
    eprint = "1503.04618",
    archivePrefix = "arXiv",
    primaryClass = "gr-qc",
    doi = "10.1088/0264-9381/32/14/145001",
    journal = "Class. Quant. Grav.",
    volume = "32",
    number = "14",
    pages = "145001",
    year = "2015"
}

@article{universe6020024,
AUTHOR = {O’Connell, Robert F.},
TITLE = {Rotation and Spin and Position Operators in Relativistic Gravity and Quantum Electrodynamics},
JOURNAL = {Universe},
VOLUME = {6},
YEAR = {2020},
NUMBER = {2},
ARTICLE-NUMBER = {24},
URL = {https://www.mdpi.com/2218-1997/6/2/24},
ISSN = {2218-1997},
DOI = {10.3390/universe6020024}
}

@article{Regge:1974zd,
    author = "Regge, Tullio and Teitelboim, Claudio",
    title = "{Role of Surface Integrals in the Hamiltonian Formulation of General Relativity}",
    reportNumber = "Print-74-0988 (IAS,PRINCETON)",
    doi = "10.1016/0003-4916(74)90404-7",
    journal = "Annals Phys.",
    volume = "88",
    pages = "286",
    year = "1974"
}

@article{Blanchet:2013haa,
    author = "Blanchet, Luc",
    title = "{Gravitational Radiation from Post-Newtonian Sources and Inspiralling Compact Binaries}",
    eprint = "1310.1528",
    archivePrefix = "arXiv",
    primaryClass = "gr-qc",
    doi = "10.12942/lrr-2014-2",
    journal = "Living Rev. Rel.",
    volume = "17",
    pages = "2",
    year = "2014"
}

@article{Porto:2016pyg,
    author = "Porto, Rafael A.",
    title = "{The effective field theorist\textquoteright{}s approach to gravitational dynamics}",
    eprint = "1601.04914",
    archivePrefix = "arXiv",
    primaryClass = "hep-th",
    doi = "10.1016/j.physrep.2016.04.003",
    journal = "Phys. Rept.",
    volume = "633",
    pages = "1--104",
    year = "2016"
}

@article{Levi:2018nxp,
    author = "Levi, Mich\`ele",
    title = "{Effective Field Theories of Post-Newtonian Gravity: A comprehensive review}",
    eprint = "1807.01699",
    archivePrefix = "arXiv",
    primaryClass = "hep-th",
    doi = "10.1088/1361-6633/ab12bc",
    journal = "Rept. Prog. Phys.",
    volume = "83",
    number = "7",
    pages = "075901",
    year = "2020"
}

@article{Levi:2014gsa,
    author = "Levi, Michele and Steinhoff, Jan",
    title = "{Leading order finite size effects with spins for inspiralling compact binaries}",
    eprint = "1410.2601",
    archivePrefix = "arXiv",
    primaryClass = "gr-qc",
    doi = "10.1007/JHEP06(2015)059",
    journal = "JHEP",
    volume = "06",
    pages = "059",
    year = "2015"
}

@article{Levi:2015msa,
    author = "Levi, Michele and Steinhoff, Jan",
    title = "{Spinning gravitating objects in the effective field theory in the post-Newtonian scheme}",
    eprint = "1501.04956",
    archivePrefix = "arXiv",
    primaryClass = "gr-qc",
    doi = "10.1007/JHEP09(2015)219",
    journal = "JHEP",
    volume = "09",
    pages = "219",
    year = "2015"
}

@article{Rothe:2010jj,
    author = "Rothe, Tilman J. and Schafer, Gerhard",
    title = "{Binary spinning black hole Hamiltonian in canonical center-of-mass and rest-frame coordinates through higher post-Newtonian order}",
    eprint = "1003.0390",
    archivePrefix = "arXiv",
    primaryClass = "gr-qc",
    doi = "10.1063/1.3448924",
    journal = "J. Math. Phys.",
    volume = "51",
    pages = "082501",
    year = "2010"
}

@article{Alessio:2023kgf,
    author = "Alessio, Francesco",
    title = "{Kerr binary dynamics from minimal coupling and double copy}",
    eprint = "2303.12784",
    archivePrefix = "arXiv",
    primaryClass = "hep-th",
    month = "3",
    year = "2023"
}

@article{Bautista:2023szu,
    author = "Bautista, Yilber Fabian",
    title = "{Dynamics for Super-Extremal Kerr Binary Systems at ${\cal O}(G^2)$}",
    eprint = "2304.04287",
    archivePrefix = "arXiv",
    primaryClass = "hep-th",
    month = "4",
    year = "2023"
}

@article{Jones:2022aji,
    author = "Jones, Callum R. T. and Solon, Mikhail",
    title = "{Scattering amplitudes and N-body post-Minkowskian Hamiltonians in general relativity and beyond}",
    eprint = "2208.02281",
    archivePrefix = "arXiv",
    primaryClass = "hep-th",
    doi = "10.1007/JHEP02(2023)105",
    journal = "JHEP",
    volume = "02",
    pages = "105",
    year = "2023"
}

@article{Pryce:1935ibt,
    author = "Pryce, M. H. L.",
    title = "{Commuting co-ordinates in the new field theory}",
    doi = "10.1098/rspa.1935.0094",
    journal = "Proc. Roy. Soc. Lond. A",
    volume = "150",
    number = "869",
    pages = "166--172",
    year = "1935"
}

@article{Pryce:1948pf,
    author = "Pryce, M. H. L.",
    title = "{The Mass center in the restricted theory of relativity and its connection with the quantum theory of elementary particles}",
    doi = "10.1098/rspa.1948.0103",
    journal = "Proc. Roy. Soc. Lond. A",
    volume = "195",
    pages = "62--81",
    year = "1948"
}

@article{Newton:1949cq,
    author = "Newton, T. D. and Wigner, Eugene P.",
    title = "{Localized States for Elementary Systems}",
    doi = "10.1103/RevModPhys.21.400",
    journal = "Rev. Mod. Phys.",
    volume = "21",
    pages = "400--406",
    year = "1949"
}

@inproceedings{Goldberger:2022ebt,
    author = "Goldberger, Walter D.",
    title = "{Effective field theories of gravity and compact binary dynamics: A Snowmass 2021 whitepaper}",
    booktitle = "{Snowmass 2021}",
    eprint = "2206.14249",
    archivePrefix = "arXiv",
    primaryClass = "hep-th",
    month = "6",
    year = "2022"
}

@article{Dlapa:2023hsl,
    author = {Dlapa, Christoph and K\"alin, Gregor and Liu, Zhengwen and Porto, Rafael A.},
    title = "{Bootstrapping the relativistic two-body problem}",
    eprint = "2304.01275",
    archivePrefix = "arXiv",
    primaryClass = "hep-th",
    reportNumber = "DESY 23-041",
    month = "4",
    year = "2023"
}

@article{Kalin:2022hph,
    author = {K\"alin, Gregor and Neef, Jakob and Porto, Rafael A.},
    title = "{Radiation-reaction in the Effective Field Theory approach to Post-Minkowskian dynamics}",
    eprint = "2207.00580",
    archivePrefix = "arXiv",
    primaryClass = "hep-th",
    reportNumber = "DESY-22-109, DESY 22-109",
    doi = "10.1007/JHEP01(2023)140",
    journal = "JHEP",
    volume = "01",
    pages = "140",
    year = "2023"
}

@article{Bini:2021gat,
    author = "Bini, Donato and Damour, Thibault and Geralico, Andrea",
    title = "{Radiative contributions to gravitational scattering}",
    eprint = "2107.08896",
    archivePrefix = "arXiv",
    primaryClass = "gr-qc",
    doi = "10.1103/PhysRevD.104.084031",
    journal = "Phys. Rev. D",
    volume = "104",
    number = "8",
    pages = "084031",
    year = "2021"
}

@article{Bini:2022enm,
    author = "Bini, Donato and Damour, Thibault and Geralico, Andrea",
    title = "{Radiated momentum and radiation reaction in gravitational two-body scattering including time-asymmetric effects}",
    eprint = "2210.07165",
    archivePrefix = "arXiv",
    primaryClass = "gr-qc",
    doi = "10.1103/PhysRevD.107.024012",
    journal = "Phys. Rev. D",
    volume = "107",
    number = "2",
    pages = "024012",
    year = "2023"
}

@article{Dlapa:2021vgp,
    author = {Dlapa, Christoph and K\"alin, Gregor and Liu, Zhengwen and Porto, Rafael A.},
    title = "{Conservative Dynamics of Binary Systems at Fourth Post-Minkowskian Order in the Large-Eccentricity Expansion}",
    eprint = "2112.11296",
    archivePrefix = "arXiv",
    primaryClass = "hep-th",
    reportNumber = "DESY 21-226",
    doi = "10.1103/PhysRevLett.128.161104",
    journal = "Phys. Rev. Lett.",
    volume = "128",
    number = "16",
    pages = "161104",
    year = "2022"
}

@article{Kalin:2020fhe,
    author = {K\"alin, Gregor and Liu, Zhengwen and Porto, Rafael A.},
    title = "{Conservative Dynamics of Binary Systems to Third Post-Minkowskian Order from the Effective Field Theory Approach}",
    eprint = "2007.04977",
    archivePrefix = "arXiv",
    primaryClass = "hep-th",
    reportNumber = "DESY 20-114, DESY-20-114, SLAC-PUB-17545",
    doi = "10.1103/PhysRevLett.125.261103",
    journal = "Phys. Rev. Lett.",
    volume = "125",
    number = "26",
    pages = "261103",
    year = "2020"
}

@article{DiVecchia:2021bdo,
    author = "Di Vecchia, Paolo and Heissenberg, Carlo and Russo, Rodolfo and Veneziano, Gabriele",
    title = "{The eikonal approach to gravitational scattering and radiation at $ \mathcal{O} $(G$^{3}$)}",
    eprint = "2104.03256",
    archivePrefix = "arXiv",
    primaryClass = "hep-th",
    reportNumber = "CERN-TH-2021-046, NORDITA 2021-028, QMUL-PH-21-17",
    doi = "10.1007/JHEP07(2021)169",
    journal = "JHEP",
    volume = "07",
    pages = "169",
    year = "2021"
}

@article{Ledvinka:2008tk,
    author = "Ledvinka, Tomas and Schaefer, Gerhard and Bicak, Jiri",
    title = "{Relativistic Closed-Form Hamiltonian for Many-Body Gravitating Systems in the Post-Minkowskian Approximation}",
    eprint = "0807.0214",
    archivePrefix = "arXiv",
    primaryClass = "gr-qc",
    doi = "10.1103/PhysRevLett.100.251101",
    journal = "Phys. Rev. Lett.",
    volume = "100",
    pages = "251101",
    year = "2008"
}

@article{Jaranowski:2012eb,
    author = "Jaranowski, Piotr and Schafer, Gerhard",
    title = "{Towards the 4th post-Newtonian Hamiltonian for two-point-mass systems}",
    eprint = "1207.5448",
    archivePrefix = "arXiv",
    primaryClass = "gr-qc",
    doi = "10.1103/PhysRevD.86.061503",
    journal = "Phys. Rev. D",
    volume = "86",
    pages = "061503",
    year = "2012"
}

@article{Cheung:2018wkq,
    author = "Cheung, Clifford and Rothstein, Ira Z. and Solon, Mikhail P.",
    title = "{From Scattering Amplitudes to Classical Potentials in the Post-Minkowskian Expansion}",
    eprint = "1808.02489",
    archivePrefix = "arXiv",
    primaryClass = "hep-th",
    reportNumber = "CALT-TH-2018-031",
    doi = "10.1103/PhysRevLett.121.251101",
    journal = "Phys. Rev. Lett.",
    volume = "121",
    number = "25",
    pages = "251101",
    year = "2018"
}

@article{Lee:2023zuu,
    author = "Lee, Hojin and Lee, Kanghoon and Lee, Sangmin",
    title = "{Poincar\'e generators at second post-Minkowskian order}",
    eprint = "2307.05626",
    archivePrefix = "arXiv",
    primaryClass = "hep-th",
    doi = "10.1007/JHEP10(2023)044",
    journal = "JHEP",
    volume = "10",
    pages = "044",
    year = "2023"
}

\end{document}